\documentclass[prc,twocolumn,nofootinbib,tightenlines,superscriptaddress,showpacs,preprintnumbers]{revtex4}

\usepackage{amsmath,amssymb,amsfonts,graphicx,subfigure,wrapfig,slashed}

\newcommand{\be}{\begin{equation}}
\newcommand{\ee}{\end{equation}}

\def\hs{\hspace}
\def\no{\nonumber}

\begin{document}

\title{Rotational states in deformed nuclei: An analytic approach}



\author{W. Bentz}
\email[Corresponding author:~]{bentz@keyaki.cc.u-tokai.ac.jp}
\affiliation{Department of Physics, School of Science, Tokai University,
             1117 Kitakaname, Hiratsuka-shi, Kanagawa 259-1292, Japan}

\author{A. Arima}
\affiliation{Musashi University,
1-26-1 Toyotama-kami, Nerima-ku, Tokyo 176-8534, Japan}

\author{J. Enders}
\affiliation{Institut f\"ur Kernphysik, Technische Universit\"at Darmstadt,
Schlossgartenstrasse 9, D-64289 Darmstadt, Germany}

\author{A. Richter}
\affiliation{Institut f\"ur Kernphysik, Technische Universit\"at Darmstadt,
Schlossgartenstrasse 9, D-64289 Darmstadt, Germany}
\affiliation{ECT$^*$, Villa Tambosi, I-38123 Villazzano (Trento), Italy}

\author{J. Wambach}
\affiliation{Institut f\"ur Kernphysik, Technische Universit\"at Darmstadt,
Schlossgartenstrasse 9, D-64289 Darmstadt, Germany}

\begin{abstract}
The consequences of the spontaneous breaking of rotational symmetry are investigated
in a field theory model for deformed nuclei, based on simple separable interactions. 
The crucial role of the Ward-Takahashi identities to describe the rotational states
is emphasized. We show explicitly 
how the rotor picture emerges from the
isoscalar Goldstone modes, and how the two-rotor model emerges from the isovector 
scissors modes. As an application of the formalism, we discuss the 
M1 sum rules in deformed nuclei, and make connection to empirical information.
\end{abstract}


\maketitle

\section{Introduction}
\setcounter{equation}{0}

The common approach to describe deformed nuclei is based on the mean field
(Hartree) approximation, where the rotational symmetry of the
Hamiltonian is spontaneously broken\cite{RO,RS}. As a consequence of the broken
symmetry, Goldstone poles emerge in the Bethe-Salpeter (BS) equation (or, 
equivalently, the RPA equation) for a particle-hole pair\cite{SAP}, similar to the case of
infinite systems\cite{NO}. For finite systems with axial symmetry,  
these intrinsic zero modes lead to the picture of collective rotation of the whole 
system around
an axis perpendicular to the symmetry axis, thereby restoring the symmetry of the original
Hamiltonian\cite{UT}. The corresponding rotational
band is characterized by {\em finite} excitation energies. The self consistency relations
for the deformed mean fields\cite{IU1} provide the necessary conditions for the
existence of the Goldstone poles and the ground state rotational band.
These features, which will be elucidated in this paper by using a simple field theory model
for nucleons, were discussed transparently in models based on bosonic degrees of freedom\cite{FU},
and form the basis for the construction of effective theories 
for deformed nuclei\cite{TP}. 

Besides these collective rotations of the whole system, which can be called
isoscalar rotations, there exist also 
rotational modes of isovector character, i.e., 
rotational vibrations of protons against neutrons\cite{SC,ZA}. 
These so-called scissors modes, which were predicted originally in the 
2-rotor model\cite{TWOROT} and further investigated by using sum rule 
methods\cite{SRM} and the Interacting Boson Model\cite{IBM}, can be excited by the 
isovector orbital part of 
the M1 operator, and decouple automatically from the 
isoscalar rotational modes if the self consistency relations are 
satisfied \cite{IU1,IU2}.
Recent experimental investigations\cite{PRC71,PRC59}
on the scissors modes have concentrated
on magnetic sum rules, which are very important because they provide the
connection of the collective modes to quantities like the effective
orbital g-factors or moments of inertia, which can be determined by
other independent analyzes\cite{LS}. The experimental and theoretical works on the scissors
modes and other magnetic dipole modes at higher excitations energies\cite{ZMS} are summarized
in a recent extensive review\cite{SCREV}.

The purpose of the present paper is threefold: First, we wish to elucidate
the importance of the Ward-Takahashi identities\cite{YT} to describe the 
intrinsic isoscalar Goldstone modes and the ground state rotational band. 
In particular, we wish to show that the results 
derived for example in Ref.\cite{IU1} can be obtained rather elegantly 
by using the Ward-Takahashi identities 
without explicit reference to single particle wave functions 
or truncations of the model space.   
Second, we wish to discuss how the isovector scissors mode, which
corresponds to a RPA solution with finite energy, leads to the two-rotor
picture in a simple model calculation. (To the best of our knowledge, 
a simple and direct derivation
of the rotor picture from the intrinsic Goldstone modes, and of the two-rotor
picture from the isovector scissors modes, has not yet been presented.) 
Third, we wish to
investigate the inverse energy weighted and energy weighted M1 sum rules, and 
discuss the relation to recent experimental
works on the scissors modes. For these purposes, we will use a simple 
field theory model based on a separable quadrupole-quadrupole ($QQ$)
interaction in the BS (RPA) framework, and consider only the orbital motion of 
the nucleons. Our main interest here is the physics of the rotational modes at 
low energies in well deformed heavy nuclei, and there is evidence both 
experimentally\cite{SCREV} and theoretically\cite{ZA} that these low energy modes are
basically of orbital character.

It should be noted here that separable 
interactions have widely been used to investigate deformed nuclei in the RPA\cite{QQ,QQ1,QQ2}.
In order to perform quantitative calculations, it is well known that pairing
plays an important role. The Nambu-Gorkov formalism \cite{NG} actually provides a simple
way to incorporate the effects of pairing into the properties of quasiparticles.  
Our purpose here, however, is to gain analytic insights into the physics
behind the rotational states and the associated sum rules in the most simple and
transparent way. Therefore, the pairing effects will not be included in the formulas, 
and numerical results will not be presented in this paper. Our hope is that the
analytic approach described here can be extended to more general
types of interactions. 

In Sect. 2 we will formulate the mean field approximation and the BS equation
for our model. In Sect. 3 we will use the Ward-Takahashi identities for
angular momentum conservation to derive several important relations and
low-energy theorems for Green functions. The connections to observables will
be established in Sect. 4, where we will discuss basic properties of 
transition matrix elements, and in Sect. 5, where the M1 sum rules will be
derived. In Sect. 6 we will discuss the derivation of the rotor picture from the isoscalar
Goldstone modes, and of the 2-rotor picture from the isovector scissors modes.   
For the derivation of the ground state rotational band, no further approximations
are necessary, but in order to derive the two-rotor picture we still find it 
necessary to assume a harmonic oscillator potential for the spherical part of the
mean field. 
A summary and an outlook are given in Sect. 7.

\section{The model}
\setcounter{equation}{0}

The Hamiltonian of the model which we will use in this paper is given by
\begin{align}
H = h_{0p} + h_{0n}
&+ \frac{\chi_{pp}}{2} \left(Q^{\dagger}_p \cdot Q_p + Q^{\dagger}_n \cdot Q_n \right) \no \\
&+ \frac{\chi_{pn}}{2} \left( Q^{\dagger}_p \cdot Q_n +  Q^{\dagger}_n \cdot Q_p \right).
\label{hqq}
\end{align}
Here ${\displaystyle h_{0 \tau} = \int {\rm d}^3 x \, \psi^{\dagger}_{\tau}(x) H_0(x) 
\psi_{\tau}(x)}$, where
${\displaystyle H_0(x) = -\Delta/2M + U_0(r)}$ with $M$ the nucleon mass and
$U_0(r)$ some spherical mean field. 
The quadrupole operator is defined by
\begin{align}
Q_{\tau}^{K} = \int {\rm d}^3 x  \, \psi^{\dagger}_{\tau}(x) Q^{K}(x) \psi_{\tau}(x) 
\,\,\,\,\,\,\,\,\,\,\,\,\,(\tau=p,n),   \label{q}
\end{align}
where $Q^{K}(x) = r^2 Y_{2 K}(\hat{x})$ \footnote{
For clarity, we will also use the
notations $Q^K_p(x)$ or $Q^K_n(x)$ to indicate whether the quadrupole
field refers to protons or neutrons. In this paper, the labels $\tau$, $\rho$,
$\lambda$ stand for protons ($p$) or neutrons ($n$). If a sum over those 
labels is involved, it will be indicated explicitly.}, 
and the products $Q^{\dagger} \cdot Q$ in
(\ref{hqq}) are defined as ${\displaystyle Q^{\dagger} \cdot Q \equiv \sum_{K=-2}^2
Q^{K \dagger} Q^K}$. 
For the coupling constants of the $QQ$ force we assume $\chi_{nn}=\chi_{pp}$ and 
$\chi_{pn}=\chi_{np}$. 
Because we treat protons and neutrons as separate particles, the interaction in (\ref{hqq})
is a mixture of pure isoscalar ($\chi_{pn}=\chi_{pp}$) and pure isovector 
($\chi_{pn}=-\chi_{pp}$) type interactions.

\subsection{Mean field approximation}

The mean field approximation is formulated 
as usual by adding and subtracting a term  
$- \sum_{\tau} \beta_{\tau} Q^0_{\tau} + C$ in (\ref{hqq}), where the parameters $\beta_{\tau}$ and
the constant $C$ will be
determined later by the requirement of self consistency. 
In order to avoid 
mathematical ambiguities in the low energy theorems to be discussed later, we also
add the terms
$ - \sum_{\tau} \varepsilon_{\tau} Q^0_{\tau}$, which explicitly break the rotational
symmetry. (In the final results the symmetry breaking parameters $\varepsilon_{\tau}$ will 
be set to zero\footnote{These symmetry breaking parameters $\varepsilon$ should 
not be confused with the single particle energies, for which we will use
the symbol $\epsilon$.}.)
In this way we obtain 
\begin{align}
H &= H_{0p} + H_{0n} + C
+ \Bigl\{\left(\beta_p-\varepsilon_p\right) \, Q_p^0 + 
\left(\beta_n-\varepsilon_n\right) \, Q_n^0 \no \\
&\hs{18mm}
+ \frac{\chi_{pp}}{2} \left(Q^{\dagger}_p \cdot Q_p +
Q^{\dagger}_n \cdot Q_n \right) \nonumber \\
&\hs{18mm}
+ \frac{\chi_{pn}}{2} 
\left( Q^{\dagger}_p \cdot
Q_n +  Q^{\dagger}_n \cdot Q_p \right) - C \Bigr\},
\label{h1}
\end{align}
where 
\begin{align}
H_{0\tau} =  \int {\rm d}^3 x \, \psi_{\tau}^{\dagger}(x) \left( - \frac{\Delta}{2M} +
U_0(x) - \beta_{\tau} Q^0(x) \right) \psi_{\tau}(x).   \label{h0}
\end{align}

We now assume that the rotational symmetry is spontaneously broken, i.e., that only the 
$K=0$ component of the quadrupole operator has a finite ground 
state expectation value:
\begin{align}
Q_{\tau}^0 = \langle Q_{\tau}^0 \rangle + : Q_{\tau}^0: \, \, ,
\label{mfa}
\end{align} 
where the dots in the second term denote normal ordering. 
We require that the part $\{\dots\}$ in (\ref{h1}) becomes a ``true'' residual  
interaction, i.e; when (\ref{mfa}) is inserted into (\ref{h1}), 
this part has neither terms linear
in $:Q_{\tau}^0$: nor constant (c-number) terms. The first requirement leads to
the self consistency relations
\begin{align}
\beta_p &= \varepsilon_p - \chi_{pp}  \langle Q_{p}^0 \rangle - 
\chi_{pn} \langle Q_{n}^0 \rangle,
\nonumber \\
\beta_n &= \varepsilon_n - \chi_{nn}  \langle Q_{n}^0 \rangle - 
\chi_{np} \langle Q_{p}^0 \rangle,
\label{gap} 
\end{align}
and the second requirement determines $C$ as
\begin{align}
C &= - \frac{\chi_{pp}}{2} \left( \langle Q_{p}^0 \rangle^2 + 
\langle Q_{n}^0 \rangle^2 \right) - \chi_{pn} \langle Q_{p}^0 \rangle \langle Q_{n}^0 \rangle.
\label{c}
\end{align}
Self consistency implies that the expectation values  $\langle Q_{\tau}^0 \rangle$
in Eqs. (\ref{gap}) themselves depend on $\beta_{\tau}$.
The Hamiltonian finally becomes 
\begin{align}
H = H_{0p} + H_{0n} + C   
+ \sum_{\tau \rho} \frac{\chi_{\tau \rho}}{2} \left(:Q^{\dagger}_{\tau}:\right) 
\cdot \left(:Q_{\rho}:\right).   \label{hfin}
\end{align}

If one employs a harmonic oscillator potential 
$U_0(r) = \left(M \tilde{\omega}^2/2\right) r^2$, 
it is
sometimes convenient to define dimensionless deformation parameters 
$\tilde{\beta}_{\tau}$ by $\beta_{\tau} = M \tilde{\omega}^2 \tilde{\beta}_{\tau}$,  
because then one can express the sum 
$U_0(x) - \beta_{\tau} Q^0(x)$ in (\ref{h0}) as a deformed harmonic oscillator potential
and apply standard methods of the Nilsson model\cite{NI}. Except for the last parts
of this paper (Sect. 6.B),  
we will keep the discussions general without specifying the form of $U_0$.

\subsection{Collective excitations}

\begin{figure}[tbp]
\begin{center}
\includegraphics[width=0.4\columnwidth]{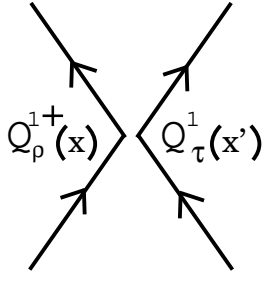}
\caption{Graphical representation of the particle-hole interaction kernel Eq.(\ref{k}).
In this and all following diagrams, time can be visualized to run from
left to right.}
\end{center}
\end{figure}

\begin{figure}[tbp]
\begin{center}
\includegraphics[width=\columnwidth]{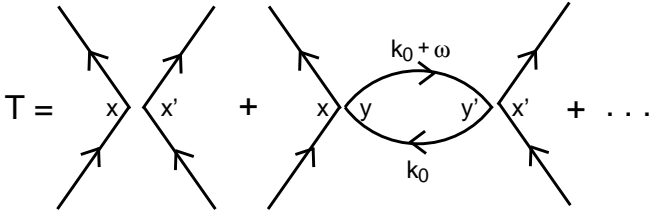}
\caption{Graphical representation of the particle-hole T-matrix Eq.(\ref{bs}) in the ladder
approximation.}
\end{center}
\end{figure}

Here we consider the Bethe-Salpeter (BS) equation, which is equivalent to the RPA
equation, for a particle and a hole in the
$K=1$ channel.
\footnote{The $K=-1$ channel is degenerate with the $K=1$ channel, while the $K=0$ 
and the $K=\pm 2$ channels have different energies.
There is no mixing of those channels for the case of axial symmetry. 
We consider the collective $K=\pm 1$ states here, because 
they correspond to the rotational states which are of main interest in this
paper.}

The residual interaction in (\ref{hfin}) is separable
in coordinate space, see Fig.1. The corresponding Feynman rule for the particle-hole 
interaction kernel is given by
\begin{align}
K_{\tau \rho}  \left({x}', {x}\right)
= - i Q^1_{\tau}({x}') \chi_{\tau \rho} Q^{1\dagger}_{\rho}({x}) . 
\label{k}
\end{align}

The inhomogeneous BS equation then reads (see Fig.2) 
\begin{align}
&T_{\tau \rho} \left({x}', {x}; \omega \right) =
K_{\tau \rho}\left({x}', {x}\right) + \int {\rm d}^3 y'
\int {\rm d}^3 y \int \frac{{\rm d}k_0}{2\pi} \sum_{\lambda} \no \\
&
K_{\tau \lambda} \left({x}', {y}'\right) 
S_{\lambda}\left({y}', {y}; k_0 + \omega\right)
S_{\lambda}\left({y}, {y}'; k_0 \right) \, T_{\lambda \rho}
\left({y}, {x}; \omega \right).
\label{bs}
\end{align}

Here we work with a mixed representation of the Feynman propagator: 
\begin{align}
S_{\tau}({x}', {x};\omega) &= \sum_{\alpha \in \tau} 
\frac{\phi_{\tau \alpha}({x}') \phi^{\dagger}_{\tau \alpha}
({x})}
{\omega - \epsilon_{\tau \alpha} + i\delta} + \sum_{i \in \tau}
\frac{\phi_{\tau i}({x}') \phi^{\dagger}_{\tau i}({x})}
{\omega - \epsilon_{\tau i} - i\delta}, \no \\
&\equiv S_{\tau P} + S_{\tau H},
\label{s}
\end{align}   
where $\phi_{\tau \alpha}({x})$ and $\epsilon_{\tau \alpha}$ are the eigenfunctions and 
eigenvalues of
$H_{0\tau}$ for particle (P) states, and $\phi_{\tau i}({x})$, $\epsilon_{\tau i}$ 
denote the corresponding quantities for hole (H) states
\footnote{Notations like
$\alpha \in \tau$ (or $i \in \tau$) indicate that the single-particle state $\alpha$ 
(or the single-hole state $i$) is a proton ($\tau=p$) or neutron ($\tau=n$) state. 
We also remark that
the states $i$ in (\ref{s}) are actually the time-reversed of the occupied single particle
states.}.

Inserting the kernel Eq.(\ref{k}) and the ansatz
\begin{align}
T_{\tau \rho}  \left({x}', {x}; \omega\right)
\equiv - i Q^1_{\tau}({x}') t_{\tau \rho}(\omega) Q^{1\dagger}_{\rho}({x})
\label{t}
\end{align}
into (\ref{bs}), we obtain the following simple matrix equation for the 
reduced T-matrix:
\begin{align}
t(\omega) &= \chi - \chi \, \pi(\omega) \, t(\omega),  \nonumber \\
\Rightarrow \quad t(\omega) &= \frac{1}{1 + \chi \pi(\omega)} \, \chi
= \chi \, \frac{1}{1 + \pi(\omega) \chi}.   \label{t1} 
\end{align}
Here the matrices in charge space have the form
\begin{align}
t = 
\begin{pmatrix}
t_{pp} & t_{pn}  \\
t_{np} & t_{nn} 
\end{pmatrix},  \quad
\chi = 
\begin{pmatrix}
\chi_{pp} & \chi_{pn}  \\
\chi_{np} & \chi_{nn}  
\end{pmatrix}, \quad
\pi =  
\begin{pmatrix}
\pi_p & 0  \\
0     & \pi_n 
\end{pmatrix}, \no \\ 
\label{mat}
\end{align}
and the proton and neutron ``bubble graphs'' $\pi_{\tau}(\omega)$ are given by 
(see Fig. 3 and Appendix A)
\begin{align}
\pi_{\tau}(\omega) &= i \int \frac{{\rm d}k_0}{2\pi} \int {\rm d}^3 x \int {\rm d}^3 x' \no \\
&\times
\left[ Q_{\tau}^{1\dagger}({x}') S_{\tau}\left({x}', {x}; k_0 + \omega \right)
Q_{\tau}^1({x}) S_{\tau}\left({x}, {x}'; k_0 \right) \right],
\nonumber \\
&= - 2 \, \sum_{(\alpha i) \in \tau} \, 
|\langle \alpha | Q^1_{\tau} | i \rangle|^2 \frac{\omega_{\alpha i}}{\omega^2
- \omega_{\alpha i}^2 + i \delta} .
\label{pi}
\end{align}
Here $\omega_{\alpha i} = \epsilon_{\alpha} - \epsilon_i$ are the non-interacting 
particle-hole energies.

\begin{figure}[tbp]
\begin{center}
\includegraphics[width=\columnwidth]{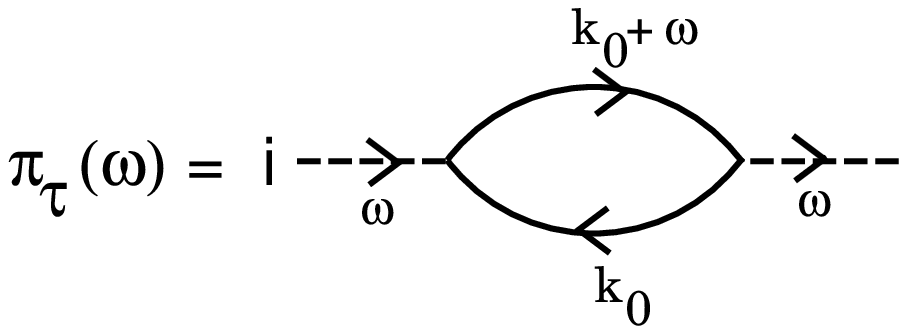}
\caption{Graphical representation of the bubble graph Eq.(\ref{pi}).}
\end{center}
\end{figure}

From (\ref{t1}), the poles of the T-matrix ($\omega^2 \equiv \omega_n^2$) are 
determined by the equation
\begin{align}
&{\rm Det} \left(1 + \chi \pi(\omega))\right)\no \\
&=
\left(1 + \pi_p(\omega) \chi_{pp}\right) \left(1 + \pi_n(\omega) \chi_{nn}\right)
- \chi_{pn}^2 \pi_p (\omega) \pi_n(\omega), \no \\
&= 0. 
\label{det}
\end{align}

\begin{figure}[ht]
\begin{center}
\includegraphics[width=\columnwidth]{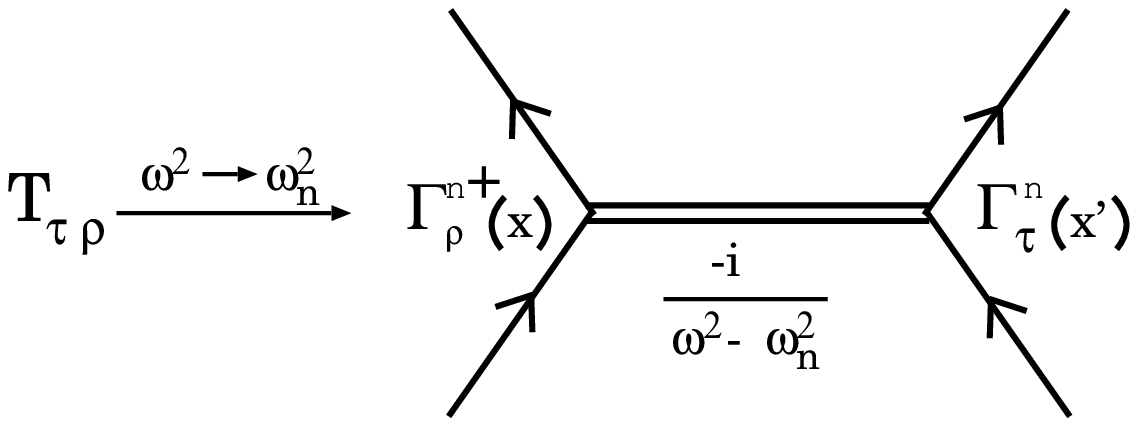}
\caption{Graphical representation of the pole behavior of the particle-hole 
T-matrix, Eq.(\ref{polef}). The double line represents a collective state.}
\end{center}
\end{figure}

It is straight forward to use Eq.(\ref{t1}) to determine the pole behavior of the
T-matrix. The results for the reduced and the full T-matrix are (see Fig. 4 for the
full T-matrix)
\begin{align}
t_{\tau \rho}(\omega) &\stackrel{\omega^2 \rightarrow \omega_n^2}{\longrightarrow}
\frac{N_{\tau}(\omega_n) N_{\rho}(\omega_n)}{\omega^2 - \omega_n^2 + i \delta}, 
\label{pole} \\
T_{\tau \rho}({x}', {x}; \omega) &\stackrel{\omega^2 \rightarrow \omega_n^2}
{\longrightarrow}
\frac{(-i) \Gamma^n_{\tau}({x'})\, \Gamma_{\rho}^{n \dagger}({x})}
{\omega^2 - \omega_n^2 + i \delta}.  
\label{polef} 
\end{align}
Here the vertex functions for the collective $K=1$ state $n$ (excitation energy
$\omega_n$) are given by
\begin{align}
\Gamma_{\tau}^n({x}) &= Q^1_{\tau}({x}) N_{\tau}(\omega_n), 
\label{gamma} \\
\Gamma_{\tau}^{n \dagger}({x}) &= Q^{1 \dagger}_{\tau}({x}) N_{\tau}(\omega_n),
\label{gammad}
\end{align}
with the normalization factors determined from\footnote{
We normalize the vertex functions $\Gamma$ as the residues at the poles in 
$\omega^2$, which corresponds to ``covariant normalization'' in
relativistic field theory.
The overall sign is chosen so that for a pure isoscalar interaction
($\chi_{pp}=\chi_{pn}$) one has $N_n/N_p = 1$, and for a pure isovector
interaction ($\chi_{pp}=-\chi_{pn}$) one has $N_n/N_p=-1$.}
\begin{align}
\frac{1}{N_p(\omega_n)^2} &= \pi_p'(\omega_n) + \pi'_n(\omega_n) \, 
\frac{N_n(\omega_n)^2}{N_p(\omega_n)^2},  \label{n1} \\
\frac{N_n(\omega_n)}{N_p(\omega_n)} &= \frac{1+\chi_{pp}\pi_p(\omega_n)}
{-\chi_{pn}\pi_n(\omega_n)}
= \frac{-\chi_{pn} \pi_p(\omega_n)}{1+\chi_{nn}\pi_n(\omega_n)}.  \label{n2}
\end{align}
Here the prime indicates differentiation w.r.t. $\omega^2$, i.e.,
\begin{align}
\pi_{\tau}'(\omega) \equiv \frac{{\rm d}\pi_{\tau}}{{\rm d}\omega^2}
= 2  \sum_{(\alpha i) \in \tau} \, 
|\langle \alpha | Q^1_{\tau} | i \rangle|^2 \frac{\omega_{\alpha i}}
{\left(\omega^2
- \omega_{\alpha i}^2 + i \delta\right)^2}.  \label{pip}
\end{align}
It is also straight forward to derive the above forms of the vertex functions
(except for the overall normalization) from the homogeneous BS equation:
Inserting the pole behavior (\ref{polef}) into Eq. 
(\ref{bs}) and taking the limit $\omega^2 \rightarrow \omega_n^2$, one obtains 
the homogeneous BS equation (see Fig. 5):
\begin{align}
&\Gamma^n_{\tau}({x}) = -i \int {\rm d}^3 y' \int{\rm d}^3 y \int \frac{{\rm d}k_0}
{2\pi}  \sum_{\rho}  \nonumber \\
&\times Q^1_{\tau}({x}) \, \chi_{\tau \rho} \, Q^{1\dagger}_{\rho}({y})
S_{\rho}({y}, {y}'; k_0+\omega) S_{\rho}({y}', {y}; k_0)\, 
\Gamma^n_{\rho}({y}').
\label{ibs}
\end{align}

\begin{figure}[tbp]
\begin{center}
\includegraphics[width=\columnwidth]{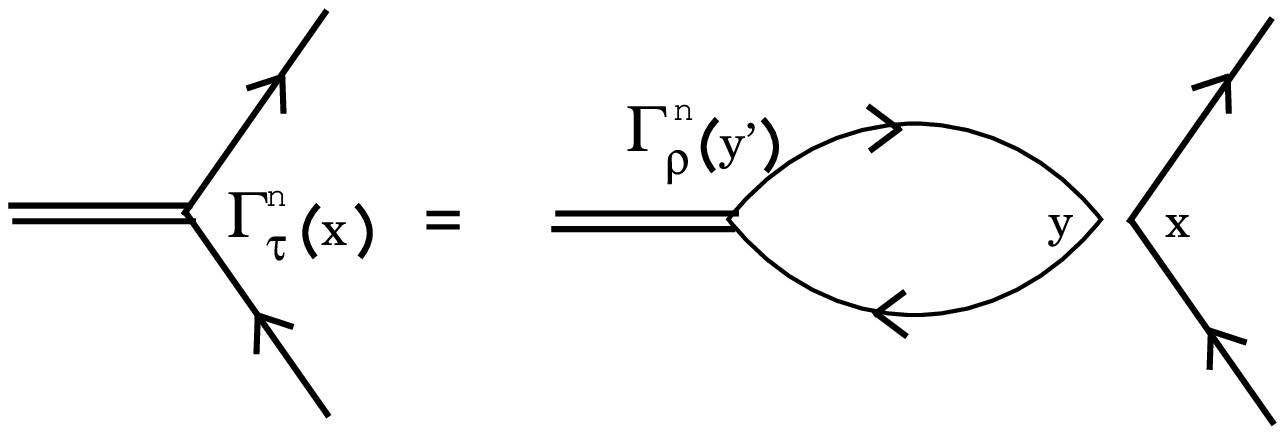}
\caption{Graphical representation of the homogeneous BS equation, Eq.(\ref{ibs}).
The double line represents a collective state.}
\end{center}
\end{figure}

Inserting here the ansatz $\Gamma^n_{\tau}({x}) = Q^1_{\tau}({x}) N_{\tau}(\omega)$,
one obtains the following matrix equation for the normalization factors:
\begin{align}
\begin{pmatrix}
N_p(\omega) \\
N_n(\omega)
\end{pmatrix}
= - 
\begin{pmatrix}
\chi_{pp} \pi_p(\omega) &  \chi_{pn} \pi_n(\omega) \\
\chi_{np} \pi_p(\omega) &  \chi_{nn}  \pi_n(\omega) 
\end{pmatrix}
\begin{pmatrix}
N_p(\omega) \\
N_n(\omega)
\end{pmatrix}.
\label{ibs1}
\end{align}
This equation again leads to the pole equation (\ref{det}) with solutions
$\omega=\omega_n$, and to the relation
(\ref{n2}).

\section{Ward-Takahashi identities}
\setcounter{equation}{0}

We first note the following commutation relation \cite{MES} between the angular
momentum operators 
${\displaystyle L^{\pm 1} \equiv \tfrac{1}{\sqrt{2}}\left(L^x \pm i L^y
\right)}$
and a tensor operator $T_{(k)}^q$ of rank $k$ with spherical components
$q=-k \dots k$ 
\footnote{The definition used here for the spherical 
components of
the angular momentum ($L^{\pm 1} = (L^x \pm i L^y)/\sqrt{2}$) differs in sign
for the $+1$ component of any other vector  
($a^1 = -(a^x + i a^y)/\sqrt{2}$,
$\,\, a^{-1} = (a^x -i  a^y)/\sqrt{2}$). Therefore, for the case 
$T_{(1)}=L$
in (\ref{comm}), we have to use $T_{(1)}^{\pm 1}
= \mp L^{\pm 1}$ to get the correct commutation relation  
$\left[L^1, L^{-1} \right] = L^0 = L^z$.}:
\begin{align}
\left[ L^{\pm 1}, T_{(k)}^q \right] = \frac{1}{\sqrt{2}}
\sqrt{k(k+1) - q (q \pm 1)} \, T_{(k)}^{q \pm 1} .   \label{comm}
\end{align}
The commutator of $H_{0\tau}$ (Eq.(\ref{h0})) with
the angular momentum operators then becomes
\begin{align}
\left[ H_{0 \tau}, L_{\tau}^{\pm 1} \right] = \sqrt{3} \beta_{\tau}
Q_{\tau}^{\pm 1},
\label{comm0}
\end{align}
and if we consider matrix elements of this identity between non-interacting 
particle-hole states, we obtain the useful relation
\begin{align}
\langle \alpha | L_{\tau}^{\pm 1} | i \rangle
= \frac{\sqrt{3} \beta_{\tau}}{\omega_{\alpha i}}
\langle \alpha | Q_{\tau}^{\pm 1} | i \rangle ,
\label{rel}
\end{align}
which will be used in later Sections. (Here 
$\left(\alpha i\right)\in \tau$.) 
   
\begin{figure*}[tbp]
\begin{center}
\includegraphics[width=2\columnwidth]{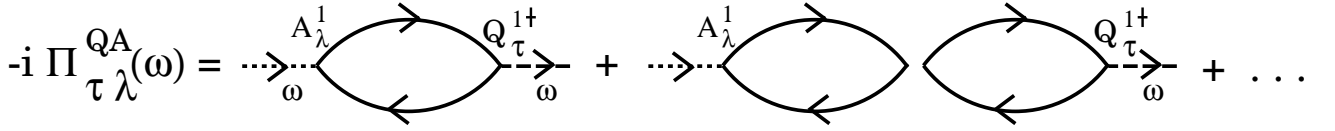}
\caption{Graphical representation of the 2-point function $\Pi^{QA}_{\tau \lambda}$, 
Eq.(\ref{rpag}).}
\end{center}
\end{figure*}  

Let us now discuss the Ward-Takahashi identity which follows from angular momentum
conservation. We consider the time derivative of the 2-point function
with external Heisenberg operators $Q^1_{\tau}(t')$ and
$L^{1}(t) = L_p^{1}(t) + L_n^1(t)$. Using the Heisenberg equation of motion
\begin{align}
\frac{\partial L^{1}}{\partial t} = i \left[ H, L^{1} \right]
= i \sqrt{3} \sum_{\tau} \varepsilon_{\tau} Q_{\tau}^{1},
\label{heis}
\end{align}
and the equal time commutator $\left[ L^1(t), Q^{1\dagger}_{\tau}(t),\right]$
from (\ref{comm}), we obtain the Ward-Takahashi identity
\begin{align}
&\frac{\partial}{\partial t} \langle 0 |
T \left( Q_{\tau}^{1 \dagger}(t') \, L^1(t) \right) |0 \rangle \nonumber \\
&= i \sqrt{3} \, \sum_{\lambda} 
\langle 0 |
T \left( Q_{\tau}^{1 \dagger}(t') \, Q_{\lambda}^1(t) \right) |0 \rangle
\, \varepsilon_{\lambda} - \sqrt{3} \delta(t-t') \langle Q_{\tau}^0  \rangle. 
\label{wt}
\end{align}
Let us define here the {\em exact} 2-point functions\footnote{We use 
the symbol $\Pi_{\lambda \tau}$ for the exact correlators and the correlators
in the chain (RPA) approximation, and $\pi_{\lambda \tau} = 
\delta_{\lambda \tau} \pi_{\tau}$ for the non-interacting ones.
Note that $\pi^{QQ}(\omega) = \pi(\omega)$ is the bubble graph of the
previous Section.} with one arbitrary
operator ($K=1$ component ${A}^1$) and the quadrupole operator 
($Q^{1 \dagger}$):  
\begin{align}
\langle 0 |
T \left( Q_{\tau}^{1 \dagger}(t') \, A^1_{\lambda}(t) \right) |0 \rangle 
&\equiv -i \, \Pi_{\tau \lambda}^{QA}(t'-t)   \no \\
&\hs{-10mm}
= -i \int {\rm d}\omega \, e^{- i \omega (t'-t)} \Pi_{\tau \lambda}^{QA}(\omega).
\label{corr} 
\end{align}
Then the Fourier transform of the Ward-Takahashi identity (\ref{wt}) 
can be expressed as
\begin{align}
\sum_{\lambda} \, \omega \, \Pi_{\tau \lambda}^{QL}(\omega) = \sqrt{3}\, 
\sum_{\lambda} \Pi_{\tau \lambda}^{QQ} (\omega) \varepsilon_{\lambda} 
- \sqrt{3} \, \langle Q^0_{\tau}  \rangle. 
\label{wt1}
\end{align}

This is the basic identity which will be used in this paper.
In the chain (RPA) approximation, the correlators
(\ref{corr}) can be expressed in terms of the reduced particle-hole t-matrix of
Eq.(\ref{t1}) as follows (see Fig. 6): 
\begin{align}
\Pi_{\tau \lambda}^{Q A}(\omega) = \delta_{\lambda \tau} 
\pi^{QA}_{\tau}(\omega) - \pi_{\tau}(\omega) t_{\tau \lambda}(\omega)
\pi^{QA}_{\lambda}(\omega) .
\label{rpag}
\end{align}
For the case $A=Q$ one can use Eq.(\ref{t1}) to simplify this expression to
\begin{align}
\Pi^{QQ}(\omega) = \frac{1}{1 + \pi(\omega) \chi} \, \pi(\omega)  ,
\label{rpa}
\end{align}
where we used the matrix notation of Eq.(\ref{mat}).

\subsection{Identities for the Goldstone modes 
($\omega\rightarrow 0$ first)}

In the limit $\omega\rightarrow 0$ (but finite $\varepsilon$), Eq.(\ref{wt1})
leads to the following low energy theorem:
\begin{align}
\sum_{\lambda} \Pi_{\tau \lambda}^{QQ} (0) \, \varepsilon_{\lambda} 
= \langle Q^0_{\tau} \rangle. \label{w}
\end{align} 
In the RPA, the correlator $\Pi^{QQ}(\omega)$ is given by (\ref{rpa}). 
Inserting this form into (\ref{w}) and multiplying from left by the
matrix $\left(1 + \pi(0) \chi \right)$ we obtain
\begin{align}
\pi(0) \left(\beta + \chi \langle Q^0 \rangle \right) = \left(1 + \pi(0) \chi
\right) \langle Q^0 \rangle ,  \nonumber
\end{align}
where we used the self consistency relation (\ref{gap}) to eliminate 
$\varepsilon$. (In this notation, 
$\beta$ and $\langle Q^0 \rangle$ are considered as vectors in charge space.)
We then obtain the identity
\begin{align}
\pi_{\tau}(0) \beta_{\tau} = \langle Q^0_{\tau} \rangle,  
\label{let}
\end{align}
which can also be shown directly by using the explicit form (\ref{pi}) of the bubble graph,
see Appendix B.

Using (\ref{let}), the self consistency relation (\ref{gap}) can be rewritten as
\begin{align}
\beta_{\tau} = \varepsilon_{\tau} - \sum_{\lambda} \left(\chi_{\tau \lambda} \pi_{\lambda}(0)\right) 
\beta_{\lambda} .  \label{gap1}
\end{align}
In the limit of exact rotational symmetry ($\varepsilon_{\tau}=0$), this relation becomes 
\begin{align}
\left(
\begin{array}{c}
\beta_p \\
\beta_n
\end{array}  \right)
= - \left(
\begin{array}{cc}
\chi_{pp} \pi_p(0) &  \chi_{pn} \pi_n(0) \\
\chi_{np}  \pi_p(0)  & \chi_{nn}   \pi_n(0) 
\end{array}  \right) 
\left(
\begin{array}{c}
\beta_p \\
\beta_n
\end{array}  \right).  \label{beta}
\end{align}
This equation leads to the condition 
${\displaystyle {\rm Det} \left(1 + \chi \pi(0)\right) = 0}$ for a 
nontrivial solution. Comparing this with the
pole equation (\ref{det}), we see that in the limit of exact rotational symmetry 
the self consistency relation guarantees the existence of a Goldstone
pole ($\omega_0=0$) in the $K=1$ channel. 

Let us now determine the vertex function for the Goldstone modes.  
From (\ref{ibs1}) and (\ref{beta}) we obtain the relation
\begin{align}
\frac{N_p(0)}{N_n(0)} = \frac{\beta_p}{\beta_n}.   \label{r}
\end{align}
Therefore the vertex function (\ref{gamma}) for the $K=1$ Goldstone mode ($n=0$) 
can be expressed as
\begin{align}
\Gamma_{\tau}^{n=0}({x}) = N \, Q^1_{\tau}({x}) \beta_{\tau} ,
\label{gammag}
\end{align}
where (see (\ref{r}) and (\ref{n1})) 
\begin{align}
N_{\tau}(0) \equiv N \beta_{\tau} , \,\,\,\,\,\,\,\,\,\,\,\,\,\,\,\,\,\,
\frac{1}{N^2} = \sum_{\tau} \beta_{\tau}^2 \pi'_{\tau}(0) .   \label{n}
\end{align}
The derivatives of the bubble graphs at $\omega=0$ are obtained from
(\ref{pip}) and (\ref{rel}) as follows:
\begin{align}  
\pi_{\tau}'(0) = 2 \sum_{(\alpha i) \in \tau} \frac{
|\langle \alpha | Q^1_{\tau} | i \rangle|^2} {\omega_{\alpha i}^3}
= \frac{2}{3 \beta_{\tau}^2} 
\sum_{(\alpha i) \in \tau} \frac{
|\langle \alpha | L^1_{\tau} | i \rangle|^2} {\omega_{\alpha i}}
\label{pip1}
\end{align}
Comparing this with the Inglis formula for the proton and neutron
moments of inertia \cite{ING}
\begin{align}
I_{\tau} = 2 \sum_{(\alpha i) \in \tau} \frac{
|\langle \alpha | L^1_{\tau} | i \rangle|^2} {\omega_{\alpha i}} , 
\label{ing}
\end{align}
we obtain the following important relations:
\begin{align}
\beta_{\tau}^2 \pi_{\tau}'(0) = \frac{I_{\tau}}{3},
\,\,\,\,\,\,\,\,\,\,\,\,\,\,\, 
N= 
\sqrt{\frac{3}{I}} , \,\,\,\,\,\,\,\,\,\,\,\,\,\,\,
N_{\tau}(0) = \sqrt{\frac{3}{I}} \, \beta_{\tau} , 
\label{i2}
\end{align}
where $I = I_p + I_n$ is the total moment of inertia.  
The Goldstone vertex function (\ref{gammag}) then takes the form
\begin{align}
\Gamma_{\tau}^{n=0}({x}) = \sqrt{\frac{3}{I}} \, Q^1_{\tau}({x}) \beta_{\tau} .
\label{gamma1}
\end{align}
The vertex function for the $K=-1$ Goldstone mode is obtained from (\ref{gamma1}) by
replacing $Q^1 \rightarrow Q^{-1}$.  
It is easy to confirm that the sum or difference of the $K=\pm 1$ vertex functions  
represents the change of the deformed mean field
($U_{\tau}(x) = U_0(x) - \beta_{\tau} Q^0_{\tau}(x)$ 
in the Hamiltonian (\ref{h0})) under an infinitesimal rotation around the $x$
or $y$ axes \cite{NO}.

\subsection{Identities for the $Q-L$ correlator ($\varepsilon\rightarrow 0$ first)}

For exact symmetry ($\varepsilon_{\tau}=0$), the Ward-Takahashi identity (\ref{wt1}) 
becomes
\begin{align}
\omega \, \sum_{\lambda} \, \Pi_{\tau \lambda}^{QL}(\omega) = - 
\sqrt{3} \, \langle Q_{\tau}^0 \rangle.
\label{wt2}
\end{align}

In the limit $\omega^2 \rightarrow \omega_n^2$, where
$\omega_n \neq 0$ is one of the nonzero solutions of the eigenvalue
equation (\ref{det}), the identity (\ref{wt2}) gives
\begin{align}
\lim_{\omega^2 \rightarrow \omega_n^2} \, 
\left(\omega^2 - \omega_n^2 \right) \sum_{\lambda} \, 
\Pi_{\tau \lambda}^{QL}(\omega) = 0  \,\,\,\,\,\,\,\,\,\,\,\,\,\,\,
(\omega_n \neq 0) .  
\label{fo}
\end{align}
Inserting here the RPA form (\ref{rpag}) and using the pole behavior of
the reduced T-matrix (\ref{pole}), we obtain
\begin{align}
\sum_{\lambda} N_{\lambda}(\omega_n) \, \pi_{\lambda}^{QL}(\omega_n) 
= 0  \,\,\,\,\,\,\,\,\,\,\,\,\,\,\,
(\omega_n \neq 0) . \label{fo1}
\end{align}
It is straight forward to check the validity of this relation
by using the explicit form of the bubble graph $\pi_{\tau}^{QL}$ (see
Appendix B). 

Next, let us consider the limit $\omega\rightarrow 0$ of (\ref{wt2}). 
For this purpose, we have to isolate the
Goldstone pole on the l.h.s. Inserting the RPA form (\ref{rpag}) of the 
correlator $\Pi^{QL}$, and isolating the Goldstone pole by using 
(\ref{pole}), i.e.,
\begin{align}
t_{\lambda' \lambda} = \frac{N_{\lambda'}(0) N_{\lambda}(0)}{\omega^2} +
\left({\rm terms}\,\,\,{\rm regular}\,\,\,{\rm for} \,\,\,\omega \rightarrow 0 
\right) ,
\label{gb}
\end{align}
the identity (\ref{wt2}) in the limit $\omega \rightarrow 0$ becomes
\begin{align}
{\rm lim}_{\omega \rightarrow 0} \sum_{\lambda}
\pi_{\tau}(0) \, N_{\tau}(0) \, N_{\lambda}(0) \,  
\left(\frac{\pi^{QL}_{\lambda}(\omega)}{\omega}\right)    = \sqrt{3} \, 
\langle Q_{\tau}^0 \rangle .
\label{ww}
\end{align}
Using here (\ref{let}) and (\ref{n}), we obtain
\begin{align}
{\rm lim}_{\omega \rightarrow 0} \, \sum_{\lambda} 
\left(\frac{\pi^{QL}_{\lambda}(\omega)}{\omega}\right) \beta_{\lambda} = \frac{I}{\sqrt{3}}.
\label{r1}
\end{align}
Again, it is easy to check this relation by using the explicit form of the bubble graph
$\pi_{\lambda}^{QL}$ and the Inglis formula, see Appendix B. Actually, because both sides
of (\ref{r1}) are one-loop quantities which consist of proton and neutron pieces,
the identity (\ref{r1}) holds for separately for protons and neutrons: 
\begin{align}
{\rm lim}_{\omega \rightarrow 0} \, 
\left(\frac{\pi^{QL}_{\lambda}(\omega)}{\omega}\right) \beta_{\lambda} = \frac{I_{\lambda}}
{\sqrt{3}} .
\label{r1a}
\end{align}
The identities (\ref{fo1}) and (\ref{r1a}) will be useful for the 
discussion of transition matrix elements in the following Section.

\section{Transition matrix elements}
\setcounter{equation}{0}

\begin{figure}[tbp]
\begin{center}
\includegraphics[width=\columnwidth]{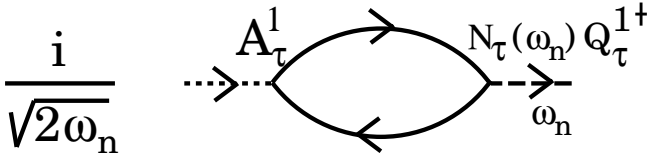}
\caption{Graphical representation of the transition matrix element, Eq.(\ref{trans}).}
\end{center}
\end{figure}

We first show that in the BS (RPA) framework the transition matrix element of the 
$K=1$ component of
any tensor operator $A^1_{\tau}$ from the ground state 
to an excited state (excitation energy $\omega_n$ and $K=1$) is given by (see Fig. 7)
\begin{align}
\langle \omega_n, K=1|{A}^1_{\tau}| 0 \rangle
= \frac{1}{\sqrt{2 \omega_n}} \pi^{QA}_{\tau}(\omega_n) \, N_{\tau}(\omega_n).   
\label{trans}
\end{align}

This formula reduces the determination of the summed transition strength 
\begin{align}
B\left(A_{\tau}; K=0 \rightarrow |K|=1\right) = 2 \sum_n 
|\langle \omega_n, K=1| {A}^1_{\tau} | 0 \rangle|^2    \label{ba}
\end{align}  
to a straight forward calculation of the Feynman diagram of Fig.7. 
(The factor $2$ comes from the contribution of $K=-1$.)

To show (\ref{trans}), we use the spectral representation of the exact correlator
(\ref{corr}): 
\begin{align}
&\Pi_{\tau \lambda}^{QA}(\omega) \no \\
&
= - \sum_n \left[ \frac{
\langle 0 | Q^{1 \dagger}_{\tau} | n \rangle \langle n | A^1_{\lambda} | 0 \rangle}
{\omega - \Omega_n + i \delta} - \frac{
\langle n | Q^{1 \dagger}_{\tau} | 0 \rangle \langle 0 | A^1_{\lambda} | n \rangle}
{\omega + \Omega_n - i \delta} \right] .
\label{spec}
\end{align}
Here $\Omega_n$ are the exact excitation energies of the eigenstates
$|n \rangle \equiv |\Omega_n; K=1 \rangle$ of the Hamiltonian $H$. 
We then obtain for $\Omega_n>0$
\begin{align}
\lim_{\omega \rightarrow \Omega_n} \left( \omega^2 - \Omega_n^2 \right)
\Pi_{\tau \lambda}^{QA}(\omega) = - 2 \Omega_n 
\langle 0 | Q^{1 \dagger}_{\tau} | n \rangle \langle n | A^1_{\lambda} | 0 \rangle .
\label{ex}
\end{align}
On the other hand, in the RPA we have from (\ref{rpag}) and the pole behavior
of the t-matrix (\ref{pole})
\begin{multline}
\lim_{\omega \rightarrow \omega_n} \left( \omega^2 - \omega_n^2 \right)
\Pi_{\tau \lambda}^{QA}(\omega) \\
= - \pi_{\tau}(\omega_n) N_{\tau}(\omega_n)
N_{\lambda}(\omega_n) \pi_{\lambda}^{QA}(\omega_n) .   
\label{nex}
\end{multline}
By comparing the r.h.s. of (\ref{ex}) in the RPA ($\Omega_n=\omega_n$) with the
r.h.s. of (\ref{nex}), 
and noting that both expressions hold also for $A=Q$ (where $\pi_{\lambda}^{QQ}
=\pi_{\lambda}$), we immediately
arrive at (\ref{trans}).

For the case $A=L= L_p + L_n$, the identity (\ref{fo1}) confirms that the total 
angular momentum
operator cannot excite a state with finite excitation energy:
\begin{align}
\langle \omega_n, K=1 | L^1 | 0 \rangle = 0  \,\,\,\,\,\,\,\,\,\,\,\,\,\,\,
(\omega_n \neq 0) .
\label{zero}
\end{align}
In fact, the state $L^1 |0 \rangle$ has zero excitation energy because of
$\left[H,L^1\right]=0$, and (\ref{zero}) confirms that this state is 
orthogonal to all states with finite excitation energy.

For the transition matrix element (\ref{trans}) of the operator $L_{\tau}$ to the 
Goldstone mode, we 
can use the form of  $N_{\tau}(0)$ from (\ref{i2}) 
and the identity (\ref{r1a}) to obtain
\begin{align}
\langle \omega_0, K=1 | L^1_{\tau} | 0 \rangle = 
\sqrt{ \frac{\omega_0}{2 I}}\, I_{\tau} 
\,\,\,\,\,\,\,\,\,\,\,\,(\omega_0 \rightarrow 0) \, .
\label{gold}
\end{align}
If we sum over protons and neutrons, we obtain an identity which follows also
directly from angular momentum conservation, Eq. (\ref{r1}):
\begin{align}
\langle \omega_0, K=1 | L^1 | 0 \rangle = 
\sqrt{ \frac{\omega_0 \, I}{2}} \,\,\,\,\,\,\,\,\,\,\,\,(\omega_0 \rightarrow 0)  .
\label{gold1}
\end{align}
This relation shows that the state $L^1| 0 \rangle$ is orthogonal to all 
RPA states, {\em including} the Goldstone mode. Actually, we will see in Sect. 6.A 
that Eq.(\ref{gold1}) is nothing but the normalization of the Goldstone state 
vectors.

\section{M1 sum rules}
\setcounter{equation}{0}

As an application of the above formalism, we consider the inverse energy weighted
(IEW) and energy weighted (EW) sum rules for the $K=1$ component of the
orbital magnetic moment (M1) operator\footnote{Generally \cite{SCREV}, 
a factor $\sqrt{3/(4\pi)}$
is included in the definition of the M1 operator. This factor is not included in our
definition.}
\begin{align}
M^1 = g_{\ell p}\, L_p^1  + g_{\ell n}\, L_n^1 ,    \label{m1}
\end{align}
where $g_{\ell \tau}$ are the orbital g-factors for $\tau=p,n$. (The free nucleon
values are $g_{\ell p}^{\rm free}=1$, $\,\,g_{\ell n}^{\rm free}=0$.)   

If we define the exact 2-point function with external M1 operators by
\begin{align}   
\Pi^{\rm MM}(\omega) &=
i \, \int {\rm d} \tau \, e^{i \omega \tau} \langle 0 | 
T \left( M^{1 \dagger}(t') \, M^1(t) \right) |0 \rangle,
\nonumber \\
&= - 2 \sum_n \, |\langle n | M^1 | 0 \rangle |^2  
\, \frac{\Omega_n}{\omega^2 - \Omega_n^2 + i \delta} ,
\label{specm}
\end{align}
where $\tau=t'-t$ and we use the notations of Eq.(\ref{spec}) for the state vectors
and energies, the  
IEW and EW sum rules can be expressed as follows\cite{LS}
\footnote{The factor 2 in these expressions counts for the contribution from
the $K=-1$ component $M^{-1}$. (Equivalently, one can express the sum rules 
by $M^x$ or $M^y$.) 
We also note that, at least in the RPA (see Eq.(\ref{gold})), the Goldstone term does not 
contribute to the spectral sum in (\ref{specm})
for finite $\omega$, and therefore also not in the $\omega \rightarrow 0$ limit 
of Eq.(\ref{iew}).} :
\begin{align}
S_{\rm IEW} &\equiv  2 \sum_{\Omega_n > 0} \frac{| \langle n| M^1 | 0\rangle |^2}{\Omega_n} 
= \lim_{\omega \rightarrow 0} \Pi^{MM}(\omega),
\label{iew}  \\
S_{\rm EW} &\equiv  2 \sum_{\Omega_n} | \langle n| M^1 | 0\rangle |^2 \, \Omega_n 
= - \lim_{\omega \rightarrow \infty} \omega^2 \, \Pi^{MM}(\omega) .
\label{ew}
\end{align}

To evaluate these sum rules in our BS (RPA) formalism, we introduce the
correlator $\Pi^{\rm LL}_{\tau \lambda}(\omega)$, which has external operators
$L^{1 \dagger}_{\tau}$ and $L^1_{\lambda}$. The form of $\Pi^{\rm LL}_{\tau \lambda}(\omega)$
in the RPA is (cf. Eq.(\ref{rpag}))
\begin{align}
\Pi^{LL}_{\tau \lambda}(\omega) = \delta_{\tau \lambda} \pi^{LL}_{\tau}
(\omega) - \pi^{LQ}_{\tau}(\omega) t_{\tau \lambda}(\omega) \pi^{QL}_{\lambda}(\omega). 
\label{rpal}
\end{align} 
Here the first term on the r.h.s. is the non-interacting bubble graph. (For the explicit form,
see Eq.(\ref{pill}).) 

Concentrating first on the IEW sum rule, we note from the Inglis
formula (\ref{ing}) that $\pi_{\tau}^{LL}(0)$ is identical to the moment of inertia:
\begin{align}
\pi_{\tau}^{LL}(\omega=0) = 2 \sum_{(\alpha i) \in \tau}
\frac{| \langle \alpha | L^1_{\tau} |i \rangle |^2}{\omega_{\alpha i}} = I_{\tau} .
\label{free}
\end{align}
To evaluate the second term in (\ref{rpal}) in the limit $\omega \rightarrow 0$, we use 
the relation (\ref{r1a}), which shows that only the first (singular) term  
in the reduced T-matrix of Eq.(\ref{gb}) contributes. Using the form of the normalization
factors given in (\ref{i2}), we obtain 
%
\begin{align}
\lim_{\omega \rightarrow 0} 
\pi^{LQ}_{\tau}(\omega) t_{\tau \lambda}(\omega) \pi^{QL}_{\lambda}(\omega)
= \frac{I_{\tau} I_{\lambda}}{I} .  \label{int}
\end{align}
The $LL$-correlator (\ref{rpal}) for $\omega \rightarrow 0$ is then obtained as
\begin{align}
\lim_{\omega \rightarrow 0} \Pi^{LL}_{\tau \lambda}(\omega) = \delta_{\tau \lambda} I_{\tau} - 
\frac{I_{\tau} I_{\lambda}}{I} .  \label{rr}
\end{align}
This result shows that the sum rule vanishes for the case where
one of the external operators is the total angular momentum $L$, i.e.,
\begin{align}
\lim_{\omega \rightarrow 0} \sum_{\lambda} \Pi^{LL}_{\tau \lambda}(\omega)=0 .
\label{cons}
\end{align} 
This is one of the many cases where the RPA-type correlations completely cancel the
non-interacting (mean field) contribution, so as to satisfy the conservation
laws (angular momentum conservation in the present case). 

Using (\ref{rr}), we obtain for the IEW sum rule (\ref{iew})
\begin{align}
S_{\rm IEW} = \lim_{\omega \rightarrow 0} 
\sum_{\tau \lambda} g_{\ell \tau} \Pi^{LL}_{\tau \lambda}
(\omega) g_{\ell \lambda} = \frac{4 I_p I_n}{I} \, \left(g_{\ell, IV}\right)^2 , 
\label{result}
\end{align}
where we define the isovector orbital g-factor by
\begin{align}
g_{\ell, IV} = 
\frac{1}{2} \left(g_{\ell p}-g_{\ell n} \right) .
\label{giv}
\end{align}

Before we continue to discuss the EW sum rule, we note the following two points: 
First, one can separate a term proportional to  
$L^1=L_p^1+L_n^1$ in the magnetic moment operator (\ref{m1}) according to
\begin{align}
M^1 = \alpha \, L^1 + (g_{\ell, p}-\alpha) \, L_p^1 + (g_{\ell, n} - \alpha) \, 
L_n^1  , \label{m1a}
\end{align}
where $\alpha$ is any number.
(The choice $\alpha=\frac{1}{2}$ leads to the conventional separation into
isoscalar and isovector parts.) From the result (\ref{cons}) it is clear that
the first term in (\ref{m1a}) does not contribute to the sum rule, 
and the contribution of the second term is independent of $\alpha$ and
given by (\ref{result}). It is possible to choose $\alpha$ so that
the RPA-type contributions vanish and the total result is given by
the non-interacting correlator. This choice is
\begin{align}
\alpha = \frac{I_p}{I} \, g_{\ell p} + \frac{I_n}{I} \, g_{\ell n} \equiv g_{\ell, IS}
,
\label{g}
\end{align}
which leads to the following separation of the magnetic moment operator
into ``isoscalar'' and ``isovector'' pieces:
\begin{align}
M^1 = g_{\ell, IS} \, L^1 + g_{\ell, IV} 
\left( \frac{2 I_n}{I} L_p^1 - \frac{2 I_p}{I} L_n^1 \right) .  
\label{split}
\end{align}
%
Both the IEW and EW sum rules discussed in this Section emerge exclusively from
the second (isovector) part of Eq.(\ref{split}). We will see in Sect. 6 that this
way to split the M1 operator follows naturally if one performs a minimal substitution
in the effective Hamiltonians for the isoscalar (Goldstone)
and isovector (scissors) rotational modes separately. 

Second, the IEW sum rule for the operator 
$(2I_n/I) L_p^1 - (2I_p/I) L_n^1$ is often interpreted as the collective 
mass parameter of the isovector rotation\cite{SRM}. The result
(\ref{result}) then confirms that this mass parameter is given by the
``isovector moment of inertia'', which is defined as
\begin{align}
I_{\rm IV} = \frac{4 I_p I_n}{I} .   \label{mass}
\end{align}

Turning now to the EW sum rule (\ref{ew}), we note that 
for the first term in the correlator (\ref{rpal}) we have
\begin{align}
- \lim_{\omega \rightarrow \infty} \omega^2 \pi_{\tau}^{LL}(\omega) &=
2 \sum_{(\alpha i) \in \tau} \omega_{\alpha i} |\langle \alpha|L^1_{\tau}|i \rangle |^2, \no \\
&= 3 \beta_{\tau}^2 \pi_{\tau}(0)  
= 3 \beta_{\tau} \langle Q_{\tau}^0 \rangle ,
\label{ew1}
\end{align}
where in the second equality we used the relation (\ref{rel}) to express the
result in terms of the bubble graph $\pi_{\tau}(0)$, and in the last equality
we used the low energy theorem (\ref{let}).  
To evaluate the second term in the RPA correlator (\ref{rpal}) in the limit
$\omega \rightarrow \infty$, we note that in this limit the t-matrix (\ref{t1}) 
becomes simply the 4-Fermi interaction constant $\chi$. Using also the form of the mixed
bubble graph $\pi_{\tau}^{LQ}(\omega)$ given in (\ref{piql}) and the identity
(\ref{rel}), we obtain
\begin{multline}
\lim_{\omega \rightarrow \infty} \omega^2 \pi_{\tau}^{LQ}(\omega) t_{\tau \lambda}(\omega)
\pi_{\lambda}^{QL}(\omega) \\
= 3 \left(\beta_{\tau}\pi_{\tau}(0)\right) \chi_{\tau \lambda}
\left(\beta_{\lambda}\pi_{\lambda}(0)\right)
= 3 \langle Q_{\tau}^0 \rangle \chi_{\tau \lambda} \langle Q_{\lambda}^0 \rangle .
\label{ew2}
\end{multline}
Adding the pieces (\ref{ew1}) and (\ref{ew2}) we obtain
\begin{multline}
- \lim_{\omega \rightarrow \infty} \omega^2 \Pi_{\tau \lambda}^{LL}(\omega) \\=
3 \left( \langle Q_{\tau}^0 \rangle \beta_{\tau} \delta_{\tau \lambda} 
+ \langle Q_{\tau}^0 \rangle \chi_{\tau \lambda} \langle Q_{\lambda}^0 \rangle \right) .
\label{ew3}
\end{multline}
Because of the self consistency relation (\ref{gap}) for exact rotational symmetry
($\varepsilon_{\tau}=0$), 
we confirm that the expression (\ref{ew3}) vanishes if we sum over $\tau$ or
$\lambda$, which is again a consequence of angular momentum conservation.
In order to get the EW sum rule for the M1 operator, we can
therefore discard the first term in (\ref{split}). By using again the relations (\ref{gap}) we finally 
obtain the following result:
\begin{align}
S_{\rm EW} &= - \lim_{\omega \rightarrow \infty} \omega^2 \sum_{\tau, \lambda} 
g_{\ell,\tau} \Pi_{\tau \lambda}^{LL}(\omega) g_{\ell, \lambda}, \no \\
&
= - 12 \left(g_{\ell, IV}\right)^2 
\langle Q_{p}^0 \rangle \chi_{pn} \langle Q_{n}^0 \rangle .
\label{ew4}
\end{align}
This is essentially the result which has been obtained in Ref.\cite{IU2} by an explicit
calculation of the corresponding double commutator. 
We also wish to mention that the EW  sum rule for the operator 
$(2I_n/I) L_p^1 - (2I_p/I) L_n^1$ is often interpreted physically in terms of the 
restoring potential energy of the isovector rotation\cite{SRM}.  
We will see in Sect.6, however, that such an intuitive result for the restoring
potential energy does not seem to emerge in our present BS (RPA) framework. 

\begin{figure}[tbp]
\begin{center}
\includegraphics[width=\columnwidth]{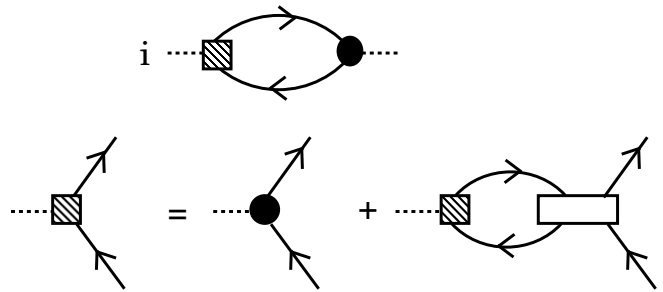}
\caption{Upper diagram: General representation of the correlator Eq.(\ref{specm}).
The black circle represents the effective irreducible particle-hole vertex, 
and the shaded square
represents the full particle-hole vertex including the RPA-type correlations. Lower diagrams:
Graphical representation of the BS equation for the full particle-hole vertex. The open square
represents the irreducible particle-hole interaction.} 
\end{center}
\end{figure}

Before we discuss the connection of these sum rules to observations, we make
a comment on the orbital g-factors:  
In a more general approach, for example the Landau-Migdal theory
\cite{MIG},
the correlator (\ref{specm}) is represented by the Feynman diagram in the first line of
Fig.8. 
The RPA-type correlations in the Landau-Migdal approach are included via
the integral equation for the total vertex, shown in the second line
of Fig.8. The driving term involves not the free but an
effective magnetic moment operator, which includes all
processes (tensor correlations, meson exchange currents, etc)
which are not taken into account by the RPA-type correlations. 
To a good approximation, this effective operator can
again be represented in the state-independent form (\ref{m1}), but now with effective
orbital g-factors, which are different from the free ones.
Therefore, in an RPA approach like our present one, the isovector g-factor
in the sum rules (\ref{result}) and (\ref{ew4}) should be considered as an 
effective quantity,
which is determined by data for magnetic moments of neighboring odd-A nuclei.
It is well known\cite{REV} that this effective isovector g-factor $g_{\ell, IV}$
is larger than the free nucleon value (which is 0.5) by approximately $10\%$.

Let us now discuss the connection to observations:
First, the $2^+$ state of the ground state rotational band is a 
classical example of an isoscalar type rotation. (Actually, in Sect.6 we 
will see how it emerges from the effective Hamiltonian for the 
isoscalar rotational mode.) Its magnetic moment is 
determined only by the first term in (\ref{split}) \cite{MGZ}, i.e.,
$\mu(2^+) = g_{\ell, IS}\, L$ with $L=2$. Its g-factor is therefore
given by
\begin{align}
g(2^+) = g_{\ell, IS} = \frac{I_p}{I} \, g_p + \frac{I_n}{I} \, g_n .
\label{gfac}
\end{align}
As a first estimate one can assume that $I_p/I \simeq Z/A$ and $I_n/I \simeq N/A$,
which gives
\begin{align}
g_{\ell, IS} \simeq \frac{Z}{A} \, g_{\ell,p} + \frac{N}{A} g_{\ell, n} = 
\frac{Z}{A},  \label{gi}
\end{align}
where the second equality has been derived rigorously from gauge invariance 
in a nuclear matter picture in Ref.\cite{BA}. 
We therefore obtain the familiar rotor value\cite{RS} 
${\displaystyle g(2^+) = \frac{Z}{A}}$. For example, the nuclei considered
in the analysis of Ref.\cite{PRC71} have $Z/A \simeq 0.4$, and the measured
g-factors are $g(2^+) \simeq 0.3$. To account for the difference,
one has to take into account the effects of pairing, which enhance the neutron 
moment of inertia relative to the proton one\cite{BM}, but qualitatively
the rotor value is correct. 

Second, it has been shown in Ref.\cite{PRC71} that the IEW sum rule value (\ref{result})
with $I_p/I \simeq Z/A$ and $I_n/I \simeq N/A$, 
agrees well with the experimentally determined ratio
$B(M1)/\omega(M1)$, where the transition matrix element $B(M1)$ and the excitation
energy $\omega(M1)$ refer to the observed low energy scissors mode. However, this
agreement is only obtained if 
the isovector orbital g-factor $g_{\ell, IV}$ is assumed to have the same value as the
isoscalar one ($g_{\ell, IS}$). As we discussed above, however, the isovector
orbital g-factor should be larger than the free nucleon value, while the isoscalar one is
smaller. This seems to indicate an interesting problem which deserves further study. 
In particular, the effects of pairing\cite{SCREV} 
should be included in the present framework. Also, for a quantitative analysis one
should extend the present model to include the effects of the nucleon spin, and 
investigate its role for the IEW sum rule, following for example the analysis of 
Ref.\cite{IU2} for the EW sum rule. The effects of pairing and spin should 
be closely investigated before drawing conclusions on the problem of the IEW sum rule.

\section{Rotational bands and scissors mode}
\setcounter{equation}{0}

In this Section we wish to derive the effective Hamiltonians and the
state vectors for the isoscalar and isovector rotational modes.
Concerning the isoscalar mode, we can apply the formalism of Sects.
2 and 3 without further approximations. For the isovector mode, however,
it seems necessary to refer to the harmonic oscillator model in order
to obtain analytic results.



Let us first establish the connection between our formalism, which is based on the
BS equation, to the usual RPA formulation in terms
of ``forward'' and ``backward'' amplitudes. This connection can easily be established
if we return to Eq.(\ref{trans}) for the transition matrix element, and  
insert the spectral form of the
non-interacting bubble graph $\pi^{QA}_{\tau}$, which is given by Eq.
(\ref{bubble}) for $B=Q$. In this way   
we obtain\footnote{We follow the notations of Ref.\cite{RO}, see in particular 
Eqs. (14.17), (14.26) and (14.32) of Ref.\cite{RO}.}
\begin{multline}
\langle \omega_n, K=1 | A^1_{\tau} | 0 \rangle \\
= \sum_{(\alpha i) \in \tau} \left[
\left(Y_{\tau}^{*}\right)_{\alpha i}(\omega_n) \, \left(A^1_{\tau}\right)_{\alpha i} +
\left(Z_{\tau}^{*}\right)_{\alpha i}(\omega_n) \, \left(A^1_{\tau}\right)_{i \alpha} \right].
\label{conv}
\end{multline}
Here we use the notation $\left({\cal O}\right)_{\alpha i} = 
\langle \alpha | {\cal O} | i \rangle$
for the particle-hole matrix elements of an operator ${\cal O}$, and 
the $K=1$ components of the RPA amplitudes are defined by
\begin{align}
\left(Y_{\tau}\right)_{\alpha i}(\omega_n) &= \frac{-1}{\sqrt{2 \omega_n}} 
\frac{ N_{\tau}(\omega_n)
\left(Q^1_{\tau}\right)_{\alpha i}} {\omega_n - \omega_{\alpha i} + i \delta},
\label{y} \\
\left(Z_{\tau}\right)_{\alpha i}(\omega_n) &= \frac{1}{\sqrt{2 \omega_n}} 
\frac{ N_{\tau}(\omega_n)
\left(Q^1_{\tau}\right)_{i \alpha}} {\omega_n + \omega_{\alpha i} - i \delta} .
\label{z} 
\end{align}
We recall two results of the
RPA: First, the states $|\omega_n, K=1 \rangle$ can be expressed by
\begin{align}
|\omega_n, K=1 \rangle = {\cal O}^{\dagger}(\omega_n, K=1) |0 \rangle , 
\label{op}
\end{align}
where ${\cal O} = \sum_{\tau} {\cal O}_{\tau}^{\dagger}$ with
%
\begin{multline}
{\cal O}_{\tau}^{\dagger}(\omega_n, K=1) \\
= \sum_{(\alpha i) \in \tau} 
\left[ \left(Y_{\tau}\right)_{\alpha i}(\omega_n) \, a^{\dagger}_{\alpha} \, a_i 
- \left(Z_{\tau}\right)_{\alpha i}(\omega_n) \, a^{\dagger}_{i} \, a_{\alpha} \right].
\label{opo}
\end{multline}
Here $a^{\dagger}$ and $a$ are the single particle creation and annihilation operators.
Second, one can derive an effective Hamiltonian for the RPA
operators ${\cal O}^{\dagger}$ and ${\cal O}$ by following bonsonization
methods based on the path integral \cite{NO}
\footnote{Bosonization methods in the path integral formalism have been used in 
relativistic field theories to show the equivalence of 4-Fermi type interactions to 
Yukawa type interactions 
\cite{BOSON1}, and also in nuclear structure physics to motivate the Interacting
Boson Model\cite{BOSON2}.}, i.e., introduce ${\cal O}^{\dagger}$, ${\cal O}$ 
as auxiliary quantities into the Hamiltonian and integrate over the
Fermion Grassmann variables. The resulting Fermionic determinant can then be expanded in
powers of ${\cal O}^{\dagger}$ and ${\cal O}$. Since here we are interested
only in the kinetic part of the effective rotational Hamiltonian ($K=\pm 1$ modes), 
we simply quote the result,
which is well known \cite{RS}\footnote{See Eq.(8.92) or (8.97) of Ref.\cite{RS}. The
constant $E_0$ in (\ref{heff0}) corresponds to $E_{HF} - \frac{1}{2} {\rm Tr} A$
in the notation of Ref.\cite{RS}.} and can be motivated also by more intuitive 
arguments:
\begin{align}
H_{\rm eff} &= E_0 + \sum_{n} H_{\rm eff}(\omega_n)\,,    \label{heff0} 
\end{align}
where
\begin{multline}
H_{\rm eff}(\omega_n) =
\frac{\omega_n}{2} \sum_{K=\pm 1} \\
\times
\left[{\cal O}^{\dagger}(\omega_n, K) {\cal O}(\omega_n, K) + {\cal O}(\omega_n, K)
{\cal O}^{\dagger}(\omega_n, K) \right].
\label{heff}
\end{multline}

\subsection{Isoscalar rotational state}

Here we derive the form of the RPA amplitudes for the Goldstone modes ($\omega_0=0$
and $K=\pm 1$),
and determine their contribution to the effective Hamiltonian (\ref{heff}). 
By using the form of $N_{\tau}(0)$ from (\ref{i2})
and the relation (\ref{rel}), we find for the amplitudes 
$Y$ and $Z$ of (\ref{y}) and (\ref{z}): 
\begin{align}
\left(Y_{\tau}\right)_{\alpha i}(\omega_0) &= \frac{\langle \alpha| L^1_{\tau}
| i \rangle}{\sqrt{2 \omega_0 I}}, \\
\left(Z_{\tau}\right)_{\alpha i}(\omega_0) &= - \frac{\langle i| L^1_{\tau}
| \alpha \rangle}{\sqrt{2 \omega_0 I}}.  \label{zg}
\end{align}
Inserting these forms into (\ref{opo}), we obtain the representation of the 
Goldstone state vector as follows:
\begin{align}
|\omega_0, K=1 \rangle &= {\cal O}^+(\omega_0, K=1)|0 \rangle , \,\,\,\,
{\rm where} \nonumber \\   
{\cal O}^{\dagger}(\omega_0, K=1) &= \frac{L^1}{\sqrt{2 \omega_0 I}} . 
\label{res1}
\end{align}
%
The creation operator for the second Goldstone mode ($K=-1$) is obtained
by the replacement $L^1 \rightarrow L^{-1}$. 

We see that the creation operator for the Goldstone mode diverges 
as $1/{\sqrt{\omega_0}}$ in the limit $\omega_0 \rightarrow 0$.
The normalization of the Goldstone state vectors, however,
is finite, and obtained from (\ref{gold1}) and (\ref{res1}) as    
\begin{align}
\sum_{K=\pm 1}   \langle \omega_0, K|\omega_0, K \rangle  = 1 . 
\label{gnorm}
\end{align} 
 
Inserting (\ref{res1}) into the effective Hamiltonian (\ref{heff}), 
we obtain for the contribution of
the Goldstone modes to the effective Hamiltonian:
\begin{align}
H_{\rm eff}(\omega_0) \equiv H_{\rm rot}(\omega_0) 
= \frac{\left(L^x \right)^2 + \left(L^y \right)^2}{2 I} = 
\frac{\vec{L}^2 - \left(L^3 \right)^2}{2 I} .
\label{rot}
\end{align}
Because this result was derived from the intrinsic Goldstone modes, which are
degenerate with the ground state of the spontaneously broken rotational symmetry, 
it corresponds to the ground state rotational band.  
The point to note is that, while the Goldstone modes have zero intrinsic excitation energy 
and correspond to
the solution $\omega_0=0$ of the BS 
equation, the operators ${\cal O}^{\dagger}(\omega_0)$ and ${\cal O}(\omega_0)$ diverge 
as $1/\sqrt{\omega_0}$, and as a result the contribution of the Goldstone modes
to the effective Hamiltonian (\ref{heff}) is finite and given by the collective rotational
energy (\ref{rot}). In a microscopic quantum theory
of finite systems, the Goldstone modes are therefore by no means ``spurious'', 
but describe the rotation of the whole 
system around an axis perpendicular to the symmetry axis, with {\em finite} rotational energy. 
Only if one {\em assumes} the rotational part (\ref{rot}) from the outset,
the Goldstone modes should be considered as ``spurious''.  

 

\subsection{Isovector rotational state (Scissors mode)}

Contrary to the isoscalar rotational modes discussed in the previous Subsection,
it seems necessary to make more specific model assumptions in order to
derive the corresponding expressions for the isovector rotational modes.
Here we will assume the harmonic oscillator (h.o.) form 
$U_0(x) = \left(M \tilde{\omega}^2/2 \right) r^2$
in the mean field Hamiltonian (\ref{h0}). The sum $U_0(x) - \beta_{\tau} Q^0(x)$
is then equivalent to a deformed harmonic oscillator potential. 
Although this model has already been used in similar contexts by many authors
\cite{IU1,IU2,SR,KS}, we discuss it in this Subsection and in Appendix C in some detail, 
because to our opinion it is highly interesting to see the isovector counterparts of
the relations given in the previous Sections, even if those are more model dependent.

In the h.o. model, the bubble graph
of Eq.(\ref{pi}) assumes a 2-pole form (see Eq.(\ref{bub1})), which makes analytic calculations
possible. Here, in order to keep the equations as schematic as possible, we
restrict ourselves to one h.o. shell ($\Delta N=0$ space), which corresponds to
the first term in Eq.(\ref{bub1}). (The case of the full h.o. space can be found in Appendix C,
and the main points will be summarized in the next Subsection.) 
In this approximation, where all particle-hole states have the same excitation 
energy ($e_{0 \tau}$),  
the bubble graph is given by the following one-pole form:
\begin{align}
\pi_{\tau}(\omega) = - 2 \frac{e_{0 \tau}}{\omega^2 - e_{0 \tau}^2 + i \delta} \,
S_{0 \tau},  \label{hpi}
\end{align}
where $e_{0 \tau}$ and $S_{0 \tau}$ are given by
%
\begin{align}
e_{0 \tau} &= K \beta_{\tau}, \qquad \qquad
\left(K = \sqrt{\frac{45}{16 \pi}} \frac{1}{M \tilde{\omega}}\right),  
\label{eps} \\
S_{0 \tau} &= \sum_0 |\langle \alpha| Q_{\tau}^1 |i \rangle|^2
= \frac{e_{0 \tau}^3}{6 \beta_{\tau}^2} \, I_{\tau}  = \frac{K^3}{6} \beta_{\tau} I_{\tau} . 
\label{s0}
\end{align}
The symbol $0$ in the sum (\ref{s0}) indicates that only the $\Delta N=0$ particle-hole states 
are included, and we used the identity (\ref{rel}) and the Inglis formula (\ref{ing}) 
to derive the second equality in
(\ref{s0}). The low energy theorem (\ref{let}) relates $S_{0 \tau}$ to
the quadrupole moment according to\footnote{As is clear from Eq.(\ref{bub1}), the 
$\Delta N=0$ sum $S_{0 \tau}$ is actually only
half of the sum in the full h.o. space. This artifact of the restriction
to one h.o. shell is formally remedied by replacing the quadrupole moment
$\langle Q_{\tau}^0 \rangle$ by the core contribution $\langle Q_{\tau}^0 \rangle_c$, which
is half of the total quadrupole moment, in all preceding relations of this paper.
See Refs.(\cite{IU1,MGZ,ZERO}) for discussions on this point.}
\begin{align}
S_{0 \tau} = \frac{K}{2} \langle Q_{\tau}^0 \rangle_c .  \label{let1}
\end{align} 
Using (\ref{eps}) and (\ref{let1}), the self consistency relations (\ref{gap}) 
for the case of exact symmetry ($\varepsilon_{\tau}=0$) can
be expressed as follows:
\begin{align}
e_{0 p} &= - 2 \chi_{pp} S_{0 p} -  2 \chi_{pn} S_{0 n},   \nonumber \\  
e_{0 n} &= - 2 \chi_{nn} S_{0 n} -  2 \chi_{np} S_{0 p}.  \label{gap0}
\end{align}
Inserting the pole form (\ref{hpi}) into the eigenvalue equation (\ref{det}), we get 
two solutions: The first one is the Goldstone solution ($\omega_0^2=0$), and the second one
is given by 
\begin{align}
\omega_1^2 &= e_{0 p}\left(e_{0 p} + 2 \chi_{pp} S_{0 p}\right) + 
e_{0 n}\left(e_{0 n} + 2 \chi_{nn} S_{0 n}\right), \nonumber \\ 
&= -2 \chi_{pn} \left (e_{0 n} S_{0 p} + e_{0 p} S_{0 n}\right)  \no \\
&= - \frac{\chi_{pn}}{3} K^4 \, \beta_p \, \beta_n \, I \,.
\label{sol}
\end{align}
Using this solution, it is easy to calculate the proton and neutron normalization
factors from Eqs.(\ref{n1}), (\ref{n2}). The result for the ratio, which corresponds 
to Eq. (\ref{r}) for the isoscalar Goldstone mode, is
\begin{align}
\frac{N_n(\omega_1)}{N_p(\omega_1)} &= \frac{\omega_1^2 - e_{0 n}^2}
{\omega_1^2 - e_{0 p}^2} \left( - \frac{S_{0 p}}{S_{0 n}} \right)
= - \frac{e_{0n} \chi_{nn} - e_{0p} \chi_{pn}}
{e_{0p} \chi_{pp} - e_{0n} \chi_{np}} . 
\label{ratio1}
\end{align}
The important point to note is that, whereas for the Goldstone mode the ratio
$N_n/N_p$ is positive and close to +1, which corresponds to an isoscalar motion, 
the ratio (\ref{ratio1}) is negative and close to -1, which corresponds 
to an isovector motion.
For the individual normalization factors we obtain (cf. Eq.(\ref{i2}) for the
Goldstone mode)
\begin{align}
N_p(\omega_1) &= - \sqrt{\frac{3}{I}} \, \beta_p \, \sqrt{\frac{I_n}{I_p}} \, 
\frac{\omega_1^2 - e_{0p}^2}{e_{0 p}^2},  \label{normp} \\ 
N_n(\omega_1) &= \sqrt{\frac{3}{I}} \, \beta_n \, \sqrt{\frac{I_p}{I_n}} \,  
\frac{\omega_1^2 - e_{0n}^2}{e_{0 n}^2}.  \label{normn} 
\end{align}
Using these normalization factors and the expression (\ref{sol}) for the
excitation energy, it is easy to calculate the RPA amplitudes from
(\ref{y}) and (\ref{z}). To illustrate the method, which is extended
to the full h.o. space in Appendix C, we note that the operator 
${\cal O}_{\tau}^{\dagger}$ of (\ref{opo}) can be expressed as follows:
\begin{align}
&{\cal O}_{\tau}^{\dagger}(\omega_n, K=1) = - \frac{N_{\tau}(\omega_n)}{\sqrt{2 \omega_n}} \no \\
&\left[ \frac{e_{0 \tau}^2}{\omega_n^2 - e_{0 \tau}^2 + i \delta} 
\frac{1}{\sqrt{3} \beta_{\tau}} \, L_{\tau}^1(0)  
+  \frac{\omega_n}{\omega_n^2 - e_{0 \tau}^2 + i \delta}  \, Q_{\tau}^1(0)
\right],
\label{q0}
\end{align}
where $n=0,\,1$. 
Here we used the identity (\ref{rel}), and defined the ``low energy part'' 
of an operator $A$ as follows:
\begin{align}
A(0) \equiv \sum_0 \left[ \left(A \right)_{\alpha i} a_{\alpha}^{\dagger} a_i 
+  \left(A \right)_{i \alpha} a_i^{\dagger} a_{\alpha} \right] .
\label{alow}
\end{align}
In the $\Delta N=0$ space, which is considered in this Subsection, we can
identify these low energy operators with the full operators ($A(0) \equiv A$).
For the Goldstone mode ($\omega_0=0$), Eq.(\ref{q0}) reproduces the general
result (\ref{res1}), and for the $\omega_1$ mode we obtain
\begin{align}
|\omega_1, K=1 \rangle &= {\cal O}^{\dagger}(\omega_1, K=1)|0 \rangle \no \\
&= \left({\cal O}^{\dagger}_L(\omega_1, K=1) + {\cal O}^{\dagger}_Q(\omega_1, K=1) \right)
|0 \rangle,
\label{lq} 
\end{align}
where the creation operator ${\cal O}^{\dagger}$ has been split into an
angular momentum part (``$L$-part'') and a quadrupole part (``$Q$-part'') defined by
\begin{align}   
&{\cal O}_L^{\dagger}(\omega_1, K=1) = \frac{1}{\sqrt{2 \omega_1 I_{IV}}} 
\left( \frac{2 I_n}{I} L_p^1 - \frac{2 I_p}{I} L_n^1 \right), 
\label{ol} \\  
&{\cal O}_Q^{\dagger}(\omega_1, K=1) \no \\
&= \frac{1}{\sqrt{2 \omega_1 I_{IV}}} 
\left( \frac{2 I_n}{I} \,  \frac{\sqrt{3} \beta_p \omega_1}{e_{0 p}^2} \, Q_p^1 
- \frac{2 I_p}{I} \, \frac{\sqrt{3} \beta_n \omega_1}{e_{0 n}^2} \, Q_n^1\right) \no \\
&= \frac{1}{\sqrt{2 \omega_1 I_{IV}}} \sqrt{\frac{- \chi_{pn} I}{\beta_p \beta_n}}
\left( \frac{2 I_n}{I} \beta_n Q_p^1 - \frac{2 I_p}{I} \beta_p Q_n^1 \right). 
\label{oq}
\end{align}
The isovector moment of inertia was defined in (\ref{mass}). 
The $L$-part (\ref{ol}) has the simple interpretation as the generator of an
out of phase rotation of protons against neutrons, where the quantities
$2I_n/I$ and $- 2I_p/I$ play the role of ``effective charges'' for protons and
neutrons, which effectively remove the
contribution of the overall in-phase rotation. The mode which is generated by
this operator is therefore called properly the ``scissors mode''.    
The $Q$-part (\ref{oq}), on the other hand, generates the quadrupole vibrations, and
we will see below that it gives
rise to the restoring force. (More general forms of those generators are given in 
Appendix C.)  

Including also the annihilation operators and the $K=-1$ mode, we can summarize
as follows:
\begin{align}
{\cal O}^{\dagger}(\omega_1, K=1) &= 
{\cal O}^{\dagger}_L(\omega_1, K=1) + {\cal O}^{\dagger}_Q(\omega_1, K=1), \nonumber \\
{\cal O}(\omega_1, K=1) &= 
{\cal O}_L(\omega_1, K=1) + {\cal O}_Q(\omega_1, K=1), \nonumber \\
{\cal O}^{\dagger}(\omega_1, K=-1) &= 
{\cal O}_L(\omega_1, K=1) - {\cal O}_Q(\omega_1, K=1), \nonumber \\
{\cal O}(\omega_1, K=-1) &= 
{\cal O}^{\dagger}_L(\omega_1, K=1) - {\cal O}^{\dagger}_Q(\omega_1, K=1) ,  
\label{list}
\end{align}
where ${\cal O}^{\dagger}_L(\omega_1, K=1)$ and ${\cal O}^{\dagger}_Q(\omega_1, K=1)$
are given in (\ref{ol}) and (\ref{oq}), and 
\begin{align}   
&{\cal O}_L(\omega_1, K=1) = \frac{1}{\sqrt{2 \omega_1 I_{IV}}} 
\left( \frac{2 I_n}{I} L_p^{-1} - \frac{2 I_p}{I} L_n^{-1} \right),
\label{ola} \\  
&{\cal O}_Q(\omega_1, K=1) =\no \\
&- \frac{1}{\sqrt{2 \omega_1 I_{IV}}} 
\sqrt{\frac{- \chi_{pn} I}{\beta_p \beta_n}}
\left( \frac{2 I_n}{I} \beta_n Q_p^{-1} - \frac{2 I_p}{I} \beta_p Q_n^{-1} \right) 
. \nonumber \\
\label{oqa}
\end{align}

By adding the $K=1$ and $K=-1$ contributions together, it is then
easy to calculate the contribution of the $\omega_1$ mode to the effective Hamiltonian 
(\ref{heff}). We obtain the following result:
\begin{align}
H_{\rm eff}(\omega_1) = H_{\rm rot}(\omega_1) + H_{Q}(\omega_1) , 
\label{hh}
\end{align}
where the rotational and quadrupole vibrational parts are given by
\begin{widetext}
\begin{align}
H_{\rm rot}(\omega_1) &=  \omega_1 \left[ {\cal O}^{\dagger}_L(\omega_1, K=1)
{\cal O}_L(\omega_1, K=1) + {\cal O}_L(\omega_1, K=1) {\cal O}^{\dagger}_L(\omega_1, K=1) \right],
\nonumber \\
&= 
\frac{1}{2 I} \left[ \frac{I_n}{I_p} 
\left( (L_p^x)^2 + (L_p^y)^2 \right) + \frac{I_p}{I_n} 
\left( (L_n^x)^2 + (L_n^y)^2 \right) 
- 2 \left(L_p^x L_n^x + L_p^y L_n^y \right) \right],
\label{hrot1} \\
H_Q(\omega_1) &= 
\omega_1 \left[ {\cal O}^{\dagger}_Q(\omega_1, K=1)
{\cal O}_Q(\omega_1, K=1) + {\cal O}_Q(\omega_1, K=1) {\cal O}^{\dagger}_Q(\omega_1, K=1) \right],
\nonumber \\
&= - \frac{\chi_{pn} I^2}{4 \beta_p \beta_n I_p I_n}
\left( \frac{2 I_n}{I} \beta_n Q_p^1 - \frac{2 I_p}{I} \beta_p Q_n^1 \right)^{\dagger}
\left( \frac{2 I_n}{I} \beta_n Q_p^1 - \frac{2 I_p}{I} \beta_p Q_n^1 \right).
\label{hq}
\end{align}
\end{widetext}
By adding (\ref{rot}) and (\ref{hh}), we obtain the total contribution of the isoscalar
and isovector rotational states to the effective Hamiltonian:
\begin{align}
H_{\rm eff}(\omega_0) +  H_{\rm eff}(\omega_1)  = H_{\rm rot} + H_Q ,  
\label{total}
\end{align}
where $H_Q=H_Q(\omega_1)$ is given by (\ref{hq}), and $H_{\rm rot}$ by the sum of
(\ref{rot}) and (\ref{hrot1}): 
\begin{align}
H_{\rm rot} &= H_{\rm rot}(\omega_0) + H_{\rm rot}(\omega_1) \no \\
&= \frac{  (L_p^x)^2 + (L_p^y)^2 }{2 I_p} + \frac{  (L_n^x)^2 + (L_n^y)^2 }{2 I_n}. 
\label{twor}
\end{align}
This is the kinetic part of the 2-rotor model Hamiltonian \cite{TWOROT}. 
We therefore obtain the important
result that the 2-rotor model is obtained from the RPA in a natural way by
adding the effective Hamiltonians for the Goldstone modes (isoscalar rotation)
and the scissors modes (isovector rotation).

We note that the two parts of the effective rotational Hamiltonian, given
by the isoscalar Goldstone part (\ref{rot}) and the isovector scissors part (\ref{hrot1}), 
correspond exactly to the isoscalar and isovector parts of the M1 operator
(\ref{split}). In the notation of first quantization, this is seen most easily 
by making a minimal substitution, namely $\vec{p}_i \rightarrow \vec{p}_i - g_{\ell p}
\vec{A}(\vec{r}_i)$ for protons ($i = 1, \dots Z$) and  
$\vec{p}_j \rightarrow \vec{p}_j - g_{\ell n}
\vec{A}(\vec{r}_j)$ for neutrons ($j = 1, \dots N$), in 
(\ref{rot}) and (\ref{hrot1}) separately.
Using the form    
$\vec{A}(\vec{r}_k) = \frac{1}{2} \left( \vec{B} \times \vec{r}_k \right)$, which
corresponds to a constant external magnetic field, and expressing the magnetic
interaction Hamiltonian in the form $- \vec{B} \cdot \vec{M}/(2M) \equiv
- \left(B^x M^x + B^y M^y \right)/(2M)$, gives the two terms in (\ref{split}). 
Viewed in this way, the presence of the scissors part (\ref{hrot1}) is
necessary to give the correct coupling to an external magnetic field. 

The quadrupole vibrational part (\ref{hq}) of the effective Hamiltonian is positive definite
(note that $\chi_{pn}<0$), and represents the restoring force which acts against
the proton-neutron oscillations in the present model. We note, however, that it
cannot be reduced to a simple geometric form, which is usually assumed in the
2-rotor model.

\subsection{Discussions}

In the previous Subsection, we have seen how the low-energy isovector scissors modes emerge 
in the simple approximation of one major h.o. shell ($\Delta N=0$).
The case of the full h.o. space is discussed in detail in Appendix C, and we can
summarize the results as follows:
There are four solutions of the RPA equation, two at low energy 
corresponding to the $\Delta N=0$ case discussed above, and two at 
high energy ($\Delta N =2$). Each of these
four modes can be represented similar to Eq.(\ref{lq}) by an $L$-part and a $Q$-part,
as shown in Eq.(\ref{q02}).  
However, the $L$-part now includes also the generator of quadrupole deformations,
describing irrotational flow, 
in addition to the generator of rotations, as shown by Eqs.(\ref{l1low}) and  
(\ref{l1high}), and the vibrational $Q$-part includes also the quadrupole operator in 
$p$-space in addition to the ordinary one in $r$-space, as shown by Eqs.(\ref{q1low}) 
and (\ref{q1high}). Concerning the $L$-part,   
for not too large deformations, the ordinary rotational term
is dominant for the low-energy solutions, while the irrotational term is dominant
for the high-energy solutions, as we explain in the last paragraph of Appendix C.

Further information on the nature of these four collective
states can be obtained by considering their vertex functions and M1 and E2 transition 
matrix elements\cite{IU2}, 
and one arrives at the following picture: 
One of the high-energy solutions ($\omega_2$ in Appendix C) carries zero M1 strength,
and is mainly of irrotational character. It can be identified as the $K=1$ component of the
isoscalar giant quadrupole resonance. The other high-energy solution ($\omega_3$) carries both 
M1 and E2 strength, and is also mainly of irrotational character. It is usually
called the the high-energy scissors mode, or equivalently the $K=1$ component of the isovector
giant quadrupole resonance\cite{SCREV}. 
The nature of the low-energy solutions is essentially the same as
we discussed in the previous Subsections, namely one is the isoscalar rotational
mode ($\omega_0$), and the other is the low-energy scissors mode ($\omega_1$), which is
mainly of rotational character.

It is very interesting to note that this picture of rotational flow at
low energy and irrotational flow at high energy is valid also for other many-body
systems, like deformed metallic clusters \cite{DMC}, deformed quantum dots \cite{DQD}, crystals
\cite{CRY}, and trapped Bose-Einstein condensates \cite{TBE}. (The recent developments
are reviewed in Ref.\cite{SCREV}.)   
For example, in the case of deformed metallic clusters,
the rotational flow corresponds to a rotation of electrons with respect to the jellium
background, and the irrotational flow to a rotation of electrons within a 
rigid surface. In the case of trapped Bose-Einstein condensates, the collective
modes are induced by an abrupt rotation of the deformed trap by a small angle, causing 
oscillations of the condensed atoms. This case is particularly 
interesting, because for normal (non-superfluid) gases one expects both the low-energy 
rotational and high-energy irrotational modes, while for the superfluid case one expects 
only the high-energy irrotational mode because of the small moment of inertia. 
Experimental evidence that the low-energy rotational mode of trapped condensed gases
indeed exists only above a critical temperature has been reported in Ref.\cite{TBE1}.

\section{Summary and outlook}
\setcounter{equation}{0}

In this paper we used a simple field theory model based on a separable $QQ$ interaction
to gain analytic insights into the physics of rotational modes in deformed nuclei.
Our essential tools were the Ward-Takahashi identities for angular momentum
conservation, which we used to discuss the Goldstone modes associated with the
spontaneous breaking of rotational symmetry, in particular their
vertex functions and decompositions into particle-hole components. In this
simple model it was possible to derive analytically the ground state rotational band
from the effective Hamiltonian for the isoscalar rotational modes. 
The isovector rotational (scissors) modes, on the other hand, 
correspond to a finite intrinsic excitation
energy, and their properties depend on the mean field and the residual proton-neutron 
interaction.
In order to obtain analytic results also for the isovector modes, we made use of the
harmonic oscillator model for the spherical part of the mean field. It was then possible
to derive the vertex functions, the decompositions into particle-hole components,
and the effective Hamiltonian also for the scissors modes. By adding the effective
Hamiltonians for the Goldstone and the scissors modes, we obtained the kinetic part of
the 2-rotor model Hamiltonian, and also the potential energy (restoring force). 
 
An other important part of our analysis was the derivation of the inverse
energy weighted and the energy weighted M1 sum rules in the RPA, where we
obtained analytic results without resorting to the harmonic oscillator
approximation. We discussed those results in connection to recent experimental
analysis on the scissors modes. We pointed out that other types of correlations,
which are not included
in the RPA (for example tensor correlations and meson exchange currents) should
enhance the M1 sum rules, because those processes are known to enhance the
isovector orbital g-factor. We pointed out an interesting
problem in this connection: The
experimental data for the inverse energy weighted sum rule seem to require that 
the isovector and isoscalar orbital g-factors are the same, while observations
and theoretical calculations of magnetic moments clearly indicate that the
isovector orbital g-factor should be larger than the isoscalar one.  
It is, however, necessary first to take into account the effects of pairing, and extend the model
to include the spin of the nucleons, before one can arrive
at firm conclusions.  

We finally mention that 
some parts of our analytic derivations should be possible for more general 
interactions, including spin-dependent and 
non-separable ones. The Ward-Takahashi identities should give an important guide
for this purpose.


\vspace{1 cm}

\noindent
{\sc Acknowledgments}  

\noindent
This work was supported through the Sonderforschungsbereich 634 of the DFG started
during an extended visit of one of the authors (W.B.) at the Institute of Nuclear
Physics of the TU Darmstadt. He furthermore expresses his thanks to Profs. P. Ring, J. Speth, 
K. Sugawara-Tanabe, P. Van Isacker, and K. Yazaki for very helpful discussions. 

\appendix

\section{Form of bubble graphs}
\setcounter{equation}{0}

\begin{figure}[tbp]
\begin{center}
\includegraphics[width=\columnwidth]{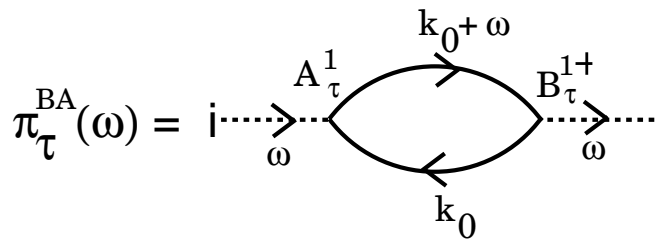}
\caption{Graphical representation of the bubble graph $\pi_{\tau}^{BA}(\omega)$.}
\end{center}
\end{figure}

In this Appendix we give the forms of various bubble graphs which appear
in the main text.

We represent a bubble graph $\pi^{BA}_{\tau}(\omega)$ with external operators 
$B^{1 \dagger}$ and $A^1$ by Fig.9. (We consider the case where these 
operators are the spherical $K=1$ components of some tensor operators, since
this is actually used in the main text.) 

Using the form (\ref{s}) of the propagators and performing 
the integration over $k_0$ by residues, we get (see Eq.(\ref{pi}) for the special
case $A=B=Q$)
\begin{multline}
\pi^{BA}_{\tau}(\omega) \\ 
= - \sum_{(\alpha i) \in \tau} 
\left[ \frac{\langle \alpha | A^{1} | i \rangle
\langle \alpha | B^1 | i \rangle ^*}{\omega - \omega_{\alpha i} + i \delta}
- \frac{\langle i | A^{1} | \alpha \rangle
\langle i | B^1 | \alpha \rangle^* }{\omega + \omega_{\alpha i} - i \delta}
\right] .
\label{bubble}
\end{multline}
In order to combine these two terms, one changes the sum over the single particle
states $(\alpha i)$ in the second term to the time reversed states $(\overline{\alpha}
\overline{i})$, which have the opposite
values of $\ell_z$ but the same energies for the axial symmetric case. Then one uses
the property (see Eq.(A.22) of \cite{RO}) 
\begin{align}
\langle \overline{i} | {\cal O} | \overline{\alpha} \rangle 
= t_{{\cal O}} \langle \alpha | {\cal O} | i \rangle ,
\label{time}
\end{align}
where $t_{\cal O} = +1$ for a T-even operator (like the quadrupole operator in our case), 
and $t_{\cal O} = -1$ for a T-odd operator (like the angular momentum operator in our
case). As a result, the relative sign of the forward and backward terms is different
for the cases where both operators have the same or the opposite T-symmetry, and one obtains
\begin{align}
\pi^{BA}_{\tau}(\omega) = - 2 \sum_{(\alpha i) \in \tau}
\langle \alpha | A^1 | i \rangle \langle \alpha | B^1 | i \rangle^* 
\frac{\Omega}{\omega^2 - \omega_{\alpha i}^2 + i \delta} ,
\label{gen}
\end{align}
where $\Omega=\omega_{\alpha i}$ if $t_A=t_B$, and $\Omega=\omega$
if $t_A = - t_B$. 

For the cases needed in the main text, the bubble graphs are then obtained as
follows:
\begin{align}
\pi^{QQ}_{\tau}(\omega) &\equiv \pi_{\tau}(\omega) =
- 2 \sum_{(\alpha i) \in \tau}
|\langle \alpha | Q^1 | i \rangle |^2  
\frac{\omega_{\alpha i}}{\omega^2 - \omega_{\alpha i}^2 + i \delta}, 
\label{piqq}  \\
\pi^{QL}_{\tau}(\omega) &=
- 2 \omega \sum_{(\alpha i) \in \tau}
\langle \alpha | L^1 | i \rangle  \langle \alpha | Q^1 | i \rangle^*   
\frac{1}{\omega^2 - \omega_{\alpha i}^2 + i \delta},  \nonumber \\
&= - 2 \omega \sum_{(\alpha i) \in \tau}
\langle \alpha | L^{-1} | i \rangle^*  \langle \alpha | Q^{-1} | i \rangle   
\frac{1}{\omega^2 - \omega_{\alpha i}^2 + i \delta},    
\label{piql}  \\
\pi^{LL}_{\tau}(\omega) &=
- 2 \sum_{(\alpha i) \in \tau}
|\langle \alpha | L^1 | i \rangle |^2  
\frac{\omega_{\alpha i}}{\omega^2 - \omega_{\alpha i}^2 + i \delta} .  
\label{pill}  
\end{align}
The two identical forms of the mixed bubble graph in (\ref{piql})
are obtained by expressing either the first or the second term in (\ref{bubble})
by the time reversed states $(\overline{\alpha} \overline{i})$. We also
note that all bubble graphs which appear in this paper correspond to
$K=1$, although this is not indicated explicitly in our notations.

\section{Identities for bubble graphs}
\setcounter{equation}{0}

In this Appendix we use the forms of the bubble graphs given in Appendix A to
confirm various identities which are derived more generally in the main
text.

\subsection{Eq.(\ref{let})}
We use Eq.(\ref{piqq}) and the identity (\ref{rel}) to write
\begin{align}
\pi_{\tau}(0) &= 2 \sum_{(PH) \in \tau} \frac{| \langle PH |Q^1 | 0 \rangle |^2}
{\omega_{PH}}  \no \\ 
&\hs{10mm}
= \frac{2}{3 \beta_{\tau}^2} \sum_{(PH) \in \tau} \omega_{PH} 
|\langle PH | L^1 | 0 \rangle |^2 ,
\label{pi0}
\end{align}
where only in this and the following equation we denote the non-interacting (mean field) ground state
simply by $|0 \rangle$, the non-interacting particle-hole states by $|PH \rangle$, and
their energies by $\omega_{PH}$. Since (\ref{pi0}) has the
form of an energy weighted sum rule, it can be expressed as a double commutator
in the standard way, and we obtain
\begin{align}
\beta_{\tau} \pi_{\tau}(0) &= - \frac{1}{3 \beta_{\tau}} 
\langle 0 | \left[ \left[ H_0, L_{\tau}^1\right], L_{\tau}^{-1} \right] | 0 \rangle, \no \\
&= - \frac{1}{\sqrt{3}} \langle 0 | \left[ Q_{\tau}^1, L_{\tau}^{^1} \right] | 0 \rangle
= \langle 0 | Q_{\tau}^0 | 0 \rangle,
\label{dc}
\end{align}
where in the last two steps we used the commutation relations (\ref{comm}) and (\ref{comm0}).

\subsection{Eq.(\ref{fo1})}

If we insert the identity (\ref{rel}) into the first form (\ref{piql}) of $\pi^{QL}_{\tau}$
and compare the result to the form (\ref{piqq}) of $\pi^{QQ}_{\tau}$ we
obtain the following identity:
\begin{align}
\omega \pi_{\tau}^{QL}(\omega) = \sqrt{3} \beta_{\tau} \left( \pi_{\tau}(\omega)
- \pi_{\tau}(0) \right) .
\label{i1}
\end{align}
Then, in order to show Eq.(\ref{fo1}), we have to show that the following relation holds
if $\omega = \omega_n$ is a solution of the eigenvalue equation (\ref{det}): 
\begin{align}
\left(\pi_p(\omega) - \pi_p(0) \right) + \frac{N_n(\omega)}{N_p(\omega)} 
\frac{\beta_n}{\beta_p}
\left(\pi_n(\omega) - \pi_n(0) \right) = 0 .  \label{one}
\end{align}
Here the ratio of the normalization factors is given by (\ref{n2}), and the
ratio of deformation parameters by (\ref{r}). Using these relations, we can
express (\ref{one}) solely by bubble graphs as follows:
\begin{align}
&\left(\pi_p(\omega) - \pi_p(0) \right) \no \\
&
+ \frac{\left(1 + \chi_{pp} \pi_p(\omega)\right)
\pi_p(0)}{\left(1 + \chi_{nn} \pi_n(0)\right) \pi_n(\omega)} 
\left(\pi_n(\omega) - \pi_n(0) \right) = 0.  \label{two}
\end{align}
More explicitly, the relation (\ref{two}) can be written as
\begin{align}
&\pi_p(\omega) \pi_n(\omega) - \pi_p(0) \pi_n(0) \no \\
&+
\chi_{pp} \Bigl[\pi_p(\omega) \pi_n(\omega) \pi_p(0) + \pi_p(\omega) \pi_n(\omega) \pi_n(0) \no \\
&\hs{10mm}
-\pi_p(\omega) \pi_p(0) \pi_n(0) - \pi_n(\omega) \pi_p(0) \pi_n(0)\Bigr] = 0.
\label{three}
\end{align}
In order to verify this relation, we use the eigenvalue equation (\ref{det}) for 
$\omega=\omega_n$ as well as for $\omega=0$. (Note that for $\omega=0$ the RPA equation
is equivalent to the self consistency relation, as we have shown in the main text.)
This gives the following identity:
\begin{align}
\frac{ \left(1 + \chi_{pp} \pi_p(\omega)\right) \left(1 + \chi_{nn} \pi_n(\omega)\right)}
{\left(1 + \chi_{pp} \pi_p(0)\right)\left(1 + \chi_{nn} \pi_n(0)\right)}
= \frac{ \pi_p(\omega) \pi_n(\omega)}{\pi_p(0) \pi_n(0)} . 
\label{four}
\end{align}
It is readily seen that this relation is the same as (\ref{three}). This concludes the
explicit verification of Eq.(\ref{fo1}). 

\subsection{Eqs.(\ref{r1}) and (\ref{r1a})} 

Using Eq.(\ref{piql}) for the bubble graph $\pi^{QL}_{\tau}$ and the identity
(\ref{rel}) we obtain
\begin{align}
\lim_{\omega \rightarrow 0} \left( \frac{\pi_{\tau}^{QL}(\omega)}{\omega}\right)
= \frac{2}{\sqrt{3} \beta_{\tau}} \sum_{(\alpha i) \in \tau}
\frac{|\langle \alpha | L^1 | i \rangle |^2}{\omega_{\alpha i}} .  \nonumber
\end{align}
Using then the Inglis formula (\ref{ing}) we obtain (\ref{r1a}).

\section{Deformed harmonic oscillator}
\setcounter{equation}{0}

In this Appendix we review some formulas 
\cite{IU1,ZA,IU2,SR,KS}   
for the RPA calculation with deformed h.o. mean fields. 

Using $U_0(x) = M \tilde{\omega}^2 r^2 /2 $ 
in the mean field Hamiltonian (\ref{h0}), the sum $U_0(x) - \beta_{\tau} Q^0(x)$
becomes a deformed h.o. potential with frequencies
\begin{align}
\tilde{\omega}_{x,\tau} &= \tilde{\omega}_{y,\tau} = \tilde{\omega}\sqrt{1 + \frac{2}{3} \delta_{\tau}}, \nonumber \\
\tilde{\omega}_{z,\tau} &= \tilde{\omega}\sqrt{1 - \frac{4}{3} \delta_{\tau}} ,  \label{fr}
\end{align}
where the dimensionless deformation parameters $\delta_{\tau}$ are related to the $\beta_{\tau}$ 
of the main text by
\begin{align}
\delta_{\tau} = \frac{K}{\tilde{\omega}} \beta_{\tau} ,
\label{delta0}
\end{align}
and $K$ is defined in Eq.(\ref{eps}).
In this simple model, two types of particle-hole excitations contribute to
the bubble graph of Eq.(\ref{pi}), corresponding to excitations
within one h.o. shell ($\Delta N=0$) and across two shells ($\Delta N=2$). 
The corresponding excitation energies are given by 
\begin{align}
e_{0 \tau} &= \tilde{\omega}_{x, \tau} - \tilde{\omega}_{z, \tau} \simeq  
\delta_{\tau} \, \tilde{\omega}\, ,   \label{e0} \\
e_{2 \tau} &=  \tilde{\omega}_{x, \tau} + \tilde{\omega}_{z, \tau} \simeq  
2 \tilde{\omega} \left(1 - \frac{1}{6} \delta_{\tau} \right). \label{e2}
\end{align}
The bubble graph (\ref{pi}) then takes the form
\begin{align}
\pi_{\tau}(\omega) &= - 2 \left( \frac{e_{0 \tau}}{\omega^2 - e_{0 \tau}^2}
S_{0 \tau} + \frac{e_{2 \tau}}{\omega^2 - e_{2 \tau}^2}
S_{2 \tau} \right), \no \\
&= 
- 2 \left( \frac{e_{0 \tau}^2}{\omega^2 - e_{0 \tau}^2}
+ \frac{e_{2 \tau}^2}{\omega^2 - e_{2 \tau}^2} \right) 
\frac{S_{0 \tau}}{e_{0 \tau}}. 
\label{bub1}
\end{align}
Here the quantities 
\begin{align}
S_{0 \tau} = \sum_0  |\langle \alpha| Q_{\tau}^1 |i \rangle|^2,
\,\,\,\,\,\,\,\,\,\,\,\,  
S_{2 \tau} = \sum_2  |\langle \alpha| Q_{\tau}^1 |i \rangle|^2,
\label{ss}
\end{align}
denote the sums over the $\Delta N=0$ and $\Delta N=2$ particle-hole states,
and in the second equality of (\ref{bub1}) we used the relation\cite{IU1}
\begin{align}
\frac{S_{0 \tau}}{e_{0 \tau}} = \frac{S_{2 \tau}}{e_{2 \tau}} .
\label{kr}
\end{align}
This relation, which follows from the analytic forms of $S_{0 \tau}$ and $S_{2 \tau}$
given by Eq.(27) of Ref.\cite{SR}, 
shows that the $\Delta N=0$ and $\Delta N=2$ excitations give
the same contributions to $\pi_{\tau}(0)$. The low energy theorem (\ref{let}) 
then can be written in the form
\begin{align}
S_{0 \tau} = \frac{K}{4} \langle Q^0_{\tau} \rangle .  
\label{let2}
\end{align}
Using (\ref{e0}) and (\ref{let2}), the self consistency relations (\ref{gap})
take the form
\begin{align}
e_{0 p} &= - 4 \chi_{pp} \, S_{0 p} -  4 \chi_{pn} \, S_{0 n},  \\  
e_{0 n} &= - 4 \chi_{nn} \, S_{0 n} -  4 \chi_{np}\, S_{0 p}.  \label{gap2}
\end{align} 
Inserting the form (\ref{bub1}) of the bubble graph into the eigenvalue equation
(\ref{det}), we can calculate the collective excitation energies in this model.
Besides the Goldstone solution ($\omega_0=0$), there are three solutions 
($\omega_1, \omega_2, \omega_3$) with
positive energy, which are determined by the following cubic equation\cite{IU1}
in $x \equiv \omega^2$: 
\begin{align}
x^3 - a x^2 + b x - c =0,   \label{cub}
\end{align}
with the coefficients
\begin{align}
a &= \left(e_{0 p}^2 + e_{2 p}^2 \right) W_p(2) + (p \rightarrow n),
\label{a} \\
b &= e_{0 p}^2 e_{2 p}^2 W_p(4) + (p \rightarrow n) \no \\
&\hs{0mm}+\frac{1}{4} \left(e_{0 p}^2 + e_{2 p}^2 \right)  \left(e_{0 n}^2 + e_{2 n}^2 \right)
\left(1 + W_p(4) + W_n(4) \right),
\label{b} \\
c &= \frac{1}{2} e_{0 p}^2 e_{2 p}^2  \left(e_{0 n}^2 + e_{2 n}^2 \right) W_p(4)
+ (p \rightarrow n) .
\label{c0}
\end{align}
Here we defined (for $k=2, 4$) 
\begin{align}
W_{\tau}(k) = 1 + k \frac{\chi_{pp}}{e_{0 \tau}} S_{0 \tau} .
\nonumber
\end{align}
For the case where the proton and neutron deformations can be assumed to
be equal ($\delta_p = \delta_n$, which implies $e_{0 p} = e_{0 n} \equiv e_0$ and
$e_{2 p} = e_{2 n} \equiv e_2$), simple analytic solutions of ({\ref{cub}) exist:
One can be obtained by noting that for
\begin{align}
\omega_2  = \sqrt{\frac{e_0^2 + e_2^2}{2}} = \sqrt{\tilde{\omega}_x^2 + \tilde{\omega}_z^2}  \label{second}
\end{align}
the bubble graph (\ref{bub1}) has the same value as for $\omega_0=0$. Since 
$\omega_0=0$ is a solution of the eigenvalue equation because of the self consistence relations, 
(\ref{second}) is also a solution.
To further understand the physical nature of this solution, we note that because of 
$\pi(\omega_2) = \pi(0)$, the identity (\ref{i1}) shows that
$\pi^{QL}(\omega_2)=0$, i.e., the M1 transition matrix element ($B(M1)$ of (\ref{trans})) 
vanishes for this mode.
On the other hand, $B(E2)$ is non-zero\cite{IU2}, and this mode can therefore be identified as the
$K=1$ component of the isoscalar giant quadrupole resonance \cite{IU2}.
The remaining two solutions can then be found by solving simple quadratic
equations. One obtains to lowest order in $\delta$:
\begin{align}
\omega_1 &= e_0 \sqrt{1 +  \frac{b}{2 + b}}, 
\label{first} \\
\omega_3 &= e_2 \sqrt{1 + \frac{b}{2}} ,   \label{third}  
\end{align}
where 
\begin{align}
b = - \frac{\chi_{pp} - \chi_{pn}}{\chi_{pp} + \chi_{pn}} 
= - \frac{\chi(T=1)}{\chi(T=0)}  \label{brat}
\end{align}
is the ratio of the isovector to the isoscalar interaction strength ($b>0$).
Analytic solutions can be worked out also for the case of different proton and 
neutron deformation parameters, although the expressions become quite long.
To summarize, there are four solutions of the RPA equation, where $\omega_0$, $\omega_1$ 
are the ``low-energy'' solutions, and $\omega_2$, $\omega_3$ the ``high-energy'' solutions.  
For each solution, one can determine the normalization factors 
$N_p(\omega_n)$ and $N_n(\omega_n)$ from (\ref{n1}) and (\ref{n2}).

Let us outline here the calculation of the creation operator (\ref{opo}) for
any of these modes, following the method explained in Refs.\cite{SR,ZA}
\footnote{For simplicity,
we write the following expressions in first quantization and omit the
distinction between protons and neutrons.}:
If we add the contribution of the $\Delta N=2$ excitations to Eq.(\ref{q0})
of the main text, we obtain
\begin{align}
{\cal O}^{\dagger}(\omega_n, K=1) &= - \frac{N(\omega_n)}{\sqrt{2 \omega_n}} \no \\
&\hs{-10mm}
\sum_{m=0,2} 
\left[ \frac{e_{m}^2}{\omega_n^2 - e_{m}^2} 
\frac{1}{\sqrt{3} \beta } \, L^1(m) 
+ \frac{\omega_n}{\omega_n^2 - e_{m}^2 } \, Q^1(m) 
\right],  \nonumber \\
&\equiv {\cal O}^{\dagger}_L (\omega_n, K=1) +
{\cal O}^{\dagger}_Q (\omega_n, K=1) .
\label{q02}
\end{align}
Here $\omega_0, \dots \omega_3$ denote the RPA eigenvalues, and in addition to the ``low energy part'' 
of an operator $A$, which was defined in (\ref{alow}) of the main text, we
also define the ``high energy part'' as
\begin{align}
A(2) = \sum_2 \left[ \left(A \right)_{\alpha i} a_{\alpha}^{\dagger} a_i 
+  \left(A \right)_{i \alpha}  a_i^{\dagger} a_{\alpha} \right] .
\label{ahigh}
\end{align}
In the harmonic oscillator model, explicit forms of the operators $L^1(m)$ 
and $Q^1(m)$, where $m=0,2$, can be derived as follows: We have
\begin{align}      
L^1 &= -i \, \sum_j \left[ \left(x + i y \right)_j p_{zj} - 
\left(p_{x} + i p_{y} \right)_j z_j \right],   \label{l1a}  \\
Q^1 &= - \sqrt{\frac{15}{8 \pi}} \sum_j \, z_j \left(x + i y \right)_j \label{q1a} , 
\end{align} 
where $j$ labels the nucleons.  
If we express (\ref{l1a}) and (\ref{q1a}) in terms of the standard creation and annihilation 
operators $a^{\dagger}_k$ and $a_k$ ($k=x,y,z$) for each particle, we obtain
two kinds of terms: The first kind involves products $a_z^{\dagger} a_i$ and $a_i^{\dagger} a_z$
with $i=x,y$, and the second kind involved products $a_z^{\dagger} a_i^{\dagger}$
and $a_z a_i$. It is clear that the first kind of operators contributes exclusively
to $\Delta N=0$ excitations (operators $A(0)$), and the second one exclusively to 
$\Delta N=2$ excitations (operators $A(2)$).
Then, for each part $A(m)$ separately, one can re-expresses the creation and 
annihilation operators by the original position and momentum operators.
In this way one obtains the decompositions $L^1 = L^1(0) + L^1(2)$ and
$Q^1 = Q^1(0) + Q^1(2)$, where
\begin{widetext}
\begin{align}
L^1(0) &= \frac{ \left(\tilde{\omega}_x + \tilde{\omega}_z \right)^2}{4 \tilde{\omega}_x \tilde{\omega}_z} 
\left \{ L^1 - i \, \frac{\tilde{\omega}_x - \tilde{\omega}_z}{\tilde{\omega}_x + \tilde{\omega}_z} 
\sum_j \left[ \left(x + i y \right)_j p_{zj} + \left(p_{x} + i p_{y}\right)_j z_j
\right] \right \}, \no \\
&= \frac{ \left(\tilde{\omega}_x + \tilde{\omega_z} \right)^2}{4 \tilde{\omega}_x \tilde{\omega}_z} \, 2i
\sum_j \left \{ \left[ r_j^{(1)} \times p_j^{(1)} \right]_{(1)}^1 + 
\frac{\tilde{\omega}_x - \tilde{\omega}_z}{\tilde{\omega}_x + \tilde{\omega}_z} \left[ r_j^{(1)} \times p_j^{(1)} 
\right]_{(2)}^1 \right \},
\label{l1low} \\
L^1(2) &= - \frac{ \left(\tilde{\omega}_x - \tilde{\omega}_z \right)^2}{4 \tilde{\omega}_x \tilde{\omega}_z} 
\left \{ L^1 - i \frac{\tilde{\omega}_x + \tilde{\omega}_z}{\tilde{\omega}_x - \tilde{\omega}_z} 
\sum_j \left[ \left(x + i y \right)_j p_{zj} + \left(p_{x} + i p_{y}\right)_j z_j
\right] \right \},  \nonumber \\
&= - \frac{ \left(\tilde{\omega}_x - \tilde{\omega}_z \right)^2}{4 \tilde{\omega}_x \tilde{\omega}_z} \, 2i 
\sum_j \left \{ \left[ r_j^{(1)} \times p_j^{(1)} \right]_{(1)}^1 + 
\frac{\tilde{\omega}_x + \tilde{\omega}_z}{\tilde{\omega}_x - \tilde{\omega}_z} \left[ r_j^{(1)} \times p_j^{(1)} 
\right]_{(2)}^1 \right \},
\label{l1high} 
\end{align}
and
\begin{align}
Q^1(0) &= \frac{1}{2} Q^1 - \sqrt{\frac{15}{8 \pi}} \frac{1}{2 M^2 \tilde{\omega}_x \tilde{\omega}_z}
\sum_j p_{z j} \left(p_{x} + i p_{y} \right)_j
= \sqrt{\frac{15}{8 \pi}} \frac{1}{2} \sum_j \left \{ \left[ r_j^{(1)} \times r_j^{(1)} 
\right]_{(2)}^1
 + \frac{1}{M^2 \tilde{\omega}_x \tilde{\omega}_z} \left[ p_j^{(1)} \times p_j^{(1)} \right]_{(2)}^1 \right \},
\label{q1low} \\
Q^1(2) &= \frac{1}{2} Q^1 + \sqrt{\frac{15}{8 \pi}} \frac{1}{2 M^2 \tilde{\omega}_x \tilde{\omega}_z}
\sum_j p_{z j} \left(p_{x} + i p_{y} \right)_j
= \sqrt{\frac{15}{8 \pi}} \frac{1}{2} \sum_j \left \{ \left[ r_j^{(1)} \times r_j^{(1)} 
\right]_{(2)}^1
 - \frac{1}{M^2 \tilde{\omega}_x \tilde{\omega}_z} \left[ p_j^{(1)} \times p_j^{(1)} \right]_{(2)}^1  \right \}.
\label{q1high}
\end{align}
\end{widetext}
Here $\left[ a^{(1)} \times b^{(1)} \right]_{(k)}^q$ denotes the tensor product, 
with rank $k$ and spherical component $q$, of two
vectors $\vec{a}$ and $\vec{b}$, according to the definitions of Ref.\cite{MES}. 
The forms given above exhaust all possible one-particle tensor operators which
can be formed from the position and momentum operators.    

Inserting these forms into (\ref{q02}), we arrive at the final expression for the
creation operator of each mode. We also note that the relations (\ref{list}) of
the main text are still
valid with the above extended operators, and therefore also
the expressions given in the first lines of Eq.(\ref{hrot1}) and (\ref{hq}) 
remain valid.

We see that, in addition to the two operators
$L^1$ and $Q^1$, the presence
of the high energy modes leads to two more types of operators, 
namely $\left[ r^{(1)} \times p^{(1)} \right]_{(2)}^1$ and 
$\left[ p^{(1)} \times p^{(1)} \right]_{(2)}^1$.  
In particular, the operator $\left[ r^{(1)} \times p^{(1)} \right]_{(2)}^1$
is the generator of quadrupole deformations\cite{ZA}, and describes irrotational 
flow. This is easily seen by noting that, for example, the part in $\{ \dots \}$ 
in the first line of Eq.(\ref{l1low}), 
which corresponds to the motion around the $x$ axis, is given by 
${\displaystyle \sum_j \left[ \left(y p_z - z p_y\right)_j + \frac{\delta}{2} 
\left(y p_z + z p_y\right)_j \right]}$. It generates the displacement
${\displaystyle \propto \left( \vec{e}_x \times \vec{r} + \frac{\delta}{2} \vec{\nabla} yz\right)}$ 
of the volume element of the liquid, which consists of a rotational and irrotational part. 
We also note that, in contrast to the angular momentum, 
the generator of quadrupole deformations is not a symmetry transformation of the Hamiltonian. 

For not too large deformation, we see from
(\ref{l1low}) and (\ref{l1high}) that the rotational term is dominant in the
low-energy part $L^1(0)$, and the irrotational term is dominant in the high-energy
part $L^1(2)$. Going back to Eq.(\ref{q02}), this implies that the low enery solutions
correspond mainly to rotational flow, and the high energy solutions mainly to
irrotational flow. Further information on the character of these modes is
obtained by considering their vertex functions and M1 and E2 transition matrix 
elements\cite{IU2}, which
leads to the following picture: The $\omega_0$ mode is the isoscalar rotational
(intrinsic Goldstone) mode, and the $\omega_2$ mode is the $K=1$ component of the
isoscalar giant quadrupole
resonance, as noted already above. The remaining modes $\omega_1$ and $\omega_3$ are of
isovector type, and carry both M1 and E2 strength.  
The $\omega_1$ mode is mainly of rotational nature and is called the 
low-energy scissors mode, while the $\omega_3$ mode is mainly of irrotational nature
and is called the high-energy scissors mode or, equivalently, the $K=1$ component
of the isovector giant quadrupole resonance\cite{SCREV}.


\end{document}